\documentclass{article}

\usepackage{microtype}
\usepackage{graphicx}
\usepackage{subfigure}
\usepackage{booktabs} 

\usepackage{multirow}
\usepackage{algorithm}
\usepackage{algorithmic}
\newcommand{\indep}{\perp \!\!\! \perp}
\usepackage{extarrows}
\usepackage{wrapfig}
\usepackage{bbding}

\usepackage{hyperref}

\usepackage[accepted]{icml2024}

\usepackage{amsmath}
\usepackage{amssymb}
\usepackage{mathtools}
\usepackage{amsthm}
\usepackage{pifont}

\usepackage[capitalize,noabbrev]{cleveref}

\theoremstyle{plain}
\newtheorem{theorem}{Theorem}[section]
\newtheorem{proposition}[theorem]{Proposition}

\theoremstyle{definition}

\newtheorem{remark}[theorem]{Remark}
\DeclareMathOperator*{\argmax}{arg\,max}
\DeclareMathOperator*{\argmin}{arg\,min}
\newcommand{\yuyi}[1]{\textcolor[rgb]{0.0,0.0,0.0}{#1}}
\usepackage[textsize=tiny]{todonotes}

\icmltitlerunning{Purify Unlearnable Examples via Rate-Constrained Variational Autoencoders}

\begin{document}

\twocolumn[
\icmltitle{Purify Unlearnable Examples via Rate-Constrained Variational Autoencoders}


\icmlsetsymbol{corresponding}{$\dagger$}

\begin{icmlauthorlist}
\icmlauthor{Yi Yu}{1,2}
\icmlauthor{Yufei Wang}{2}
\icmlauthor{Song Xia}{2}
\icmlauthor{Wenhan Yang}{corresponding,4}
\icmlauthor{Shijian Lu}{3}
\icmlauthor{Yap-Peng Tan}{2}
\icmlauthor{Alex C. Kot}{2}
\end{icmlauthorlist}

\icmlaffiliation{1}{Rapid-Rich Object Search Lab, Interdisciplinary Graduate Programme, Nanyang Technological University, Singapore}
\icmlaffiliation{2}{School of Electrical and Electronic Engineering, Nanyang Technological University, Singapore}
\icmlaffiliation{3}{School of Computer Science and Engineering, Nanyang Technological University, Singapore}
\icmlaffiliation{4}{PengCheng Laboratory, Shenzhen, China}

\icmlcorrespondingauthor{Yi Yu}{yuyi0010@e.ntu.edu.sg}
\icmlcorrespondingauthor{Wenhan Yang}{yangwh@pcl.ac.cn}

\icmlkeywords{unlearnable examples, poisoning attacks, defenses, purification, perturbation availability poisoning attacks}

\vskip 0.3in
]



\printAffiliationsAndNotice{{$^\dagger\text{Corresponding author}~$}}  

\begin{abstract}
Unlearnable examples (UEs) seek to maximize testing error by making subtle modifications to training examples that are correctly labeled.
Defenses against these poisoning attacks can be categorized based on whether specific interventions are adopted during training.
The first approach is training-time defense, such as adversarial training, which can mitigate poisoning effects but is computationally intensive.
The other approach is pre-training purification, \textit{e.g.,} image short squeezing, which consists of several simple compressions but often encounters challenges in dealing with various UEs.
Our work provides a novel disentanglement mechanism to build an efficient pre-training purification method.
Firstly, we uncover rate-constrained variational autoencoders (VAEs), demonstrating a clear tendency to suppress the perturbations in UEs. We subsequently conduct a theoretical analysis for this phenomenon.
Building upon these insights, we introduce a disentangle variational autoencoder (D-VAE), capable of disentangling the perturbations with learnable class-wise embeddings.
Based on this network, a two-stage purification approach is naturally developed. The first stage focuses on roughly eliminating perturbations, while the second stage produces refined, poison-free results, ensuring effectiveness and robustness across various scenarios.
Extensive experiments demonstrate the remarkable performance of our method across CIFAR-10, CIFAR-100, and a 100-class ImageNet-subset.
{Code is available at {\url{https://github.com/yuyi-sd/D-VAE}}.}
\end{abstract}

\begin{figure*}[t]
\centering
\includegraphics[width=0.95\linewidth]{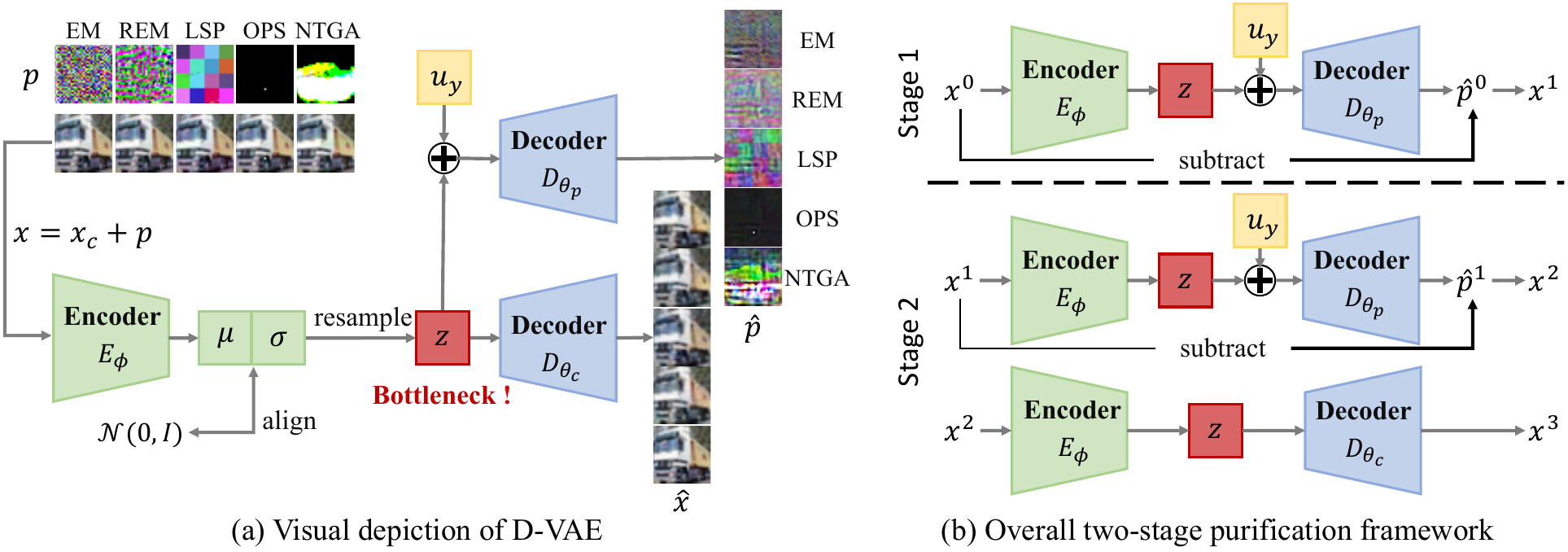}
\vspace{-4.0mm}
\caption{
(a) Visual depiction of D-VAE containing two components. One component generates reconstructed images $\boldsymbol{\hat{x}}$, preserving the primary content of unlearnable inputs $\boldsymbol{x}$. The auxiliary decoder maps a trainable class-wise embedding {$\boldsymbol{u_y}$} and latents $\boldsymbol{z}$ to disentangled perturbations $\hat{\boldsymbol{p}}$. Here, $\boldsymbol{x_c}$ is clean data, and $\boldsymbol{p}$ is added perturbations. Perturbations are normalized for better views.
(b) The purification framework consisting of two stages. The overall purification can be formulated as $\boldsymbol{x}^3 = \boldsymbol{g}(\boldsymbol{x}^0)$, where $\boldsymbol{x}^0$ is the original unlearnable data.
}
\vspace{-2mm}
\label{dvae}
\end{figure*}

\section{Introduction}
Although machine learning models often achieve impressive performance on a range of challenging tasks, their effectiveness can significantly deteriorate in the presence of the gaps between the training and testing data distributions.
%
One of the most widely studied types of these gaps is related to the vulnerability of standard models to adversarial examples~\citep{fgsm,yu2022towards,xia2024mitigating,wang2024benchmarking}, posing a significant threat to the inference phase.
However, a destructive and often underestimated threat emerges from malicious perturbations at the training phase, namely unlearnable examples, which seek to maximize testing error by making subtle modifications of correctly labeled training examples~\citep{dc}.

In the era of big data, vast amounts are freely collected from the Internet, powering advances in DNNs~\citep{schmidhuber2015deep}. Nonetheless, it's essential to note that online data may contain proprietary or private information, raising concerns about unauthorized use.
UEs are considered a promising route for data protection~\citep{em}. 
Recently, many efforts have emerged to add invisible perturbations to images as shortcuts to disrupt the training process~\citep{lsp,lin2024safeguarding}.
On the other hand, data exploiters perceive these protection techniques as potential threats to a company's commercial interests, leading to extensive research efforts in developing defenses.
Previous research has demonstrated that training-time defenses, such as adversarial training and adversarial augmentations, can alleviate poisoning effects. However, their practicality is limited by the massive computational costs.
Recently, preprocessing-based defenses have gained attention with simple compressions like JPEG and grayscale demonstrating the advantages over adversarial training in computational efficiency~\citep{iss}.
However, these methods lack universality, as different compression techniques might be best suited for various attacks.
%
Pre-training purification has demonstrated great potential in addressing the issue of UEs in both effectiveness and efficiency~\citep{iss}.
This kind of method doesn't intervene in the model's training but instead concentrates on refining the data, which well aligns with the recent theme of data-centric AI (DCAI)~\citep{zha2023data}.
%
%
Focusing on fundamental data-related issues rather than relying on untrusted or compromised data leads to more reliable and effective machine learning models.


%
In this paper, we focus on the pre-training purification paradigm. 
Our overall approach is to utilize a disentanglement mechanism to separate the poison signal from the intrinsic signal of the image with a rate-constrained VAE to obtain clean data.
Firstly, we discover that a rate-constrained VAE can effectively remove the added perturbations by constraining the KL divergence in latents when compared to JPEG~\citep{guo2018countering} in Sec~\ref{sec3.2}, with a derived detailed theoretical explanation in Sec~\ref{vae_explaination}.
Specifically, we formulate UEs as the transformation of less-predictive features into highly predictive ones. This perspective reveals that perturbations with a larger inter-class distance and smaller intra-class variance can create stronger attacks by shifting the optimal separating hyperplane of a Bayes classifier.
Subsequently, we show that VAEs are particularly effective at suppressing perturbations possessing these characteristics. Furthermore, we observe that most {perturbations} exhibit lower class-conditional entropy. Thus, we propose a method involving learnable class-wise embeddings to disentangle these added perturbations.

Building upon these findings, we present a purification framework that offers consistent and adaptable defense against UEs in Sec~\ref{sec_dvae} and Sec~\ref{sec_framework}. 
In Figure~\ref{dvae} (a), we present the D-VAE, comprising two components, capable of generating a reconstructed image $\boldsymbol{\hat{x}}$ with minimal poisoning perturbations and disentangling predicted perturbations $\boldsymbol{\hat{p}}$ with a trainable class-wise embedding $\boldsymbol{u_y}$.
Subsequently, leveraging D-VAE, we propose a two-stage purification framework illustrated in Figure~\ref{dvae} (b).
In each stage, we train D-VAE on the unlearnable dataset and perform inference using the trained D-VAE on the same dataset.
Our two-stage purification framework primarily involves two operations: 1) Estimating perturbations $\boldsymbol{\hat{p}}$ and subtracting them from $\boldsymbol{x}$;
2) Obtaining reconstructed data $\boldsymbol{\hat{x}}$ from $D_{\theta_c}$ to serve as purified images.
While the subtraction process occurs at both stages, the acquisition of $\boldsymbol{\hat{x}}$ takes place at the end of the second stage.
With this method, models trained on our purified datasets
can achieve significant boosts compared with previous SOTA methods: improved from 84\% to 90\% on CIFAR-10~\citep{cifar} and from 64\% to 75\% on the
ImageNet-subset~\citep{imagenet}.
%
%

In summary, our contributions can be outlined as follows:

$\bullet$ We discover that rate-constrained VAEs exhibit a preference for removing {perturbations} in UEs, and offer a comprehensive theoretical analysis to support this finding.

$\bullet$ We introduce D-VAE, a network that can disentangle the added perturbations and generate purified data. Our additional evaluations also show that D-VAE can purify UEs from a mixed dataset, and is able to produce new UEs, even if it only accesses to just a small fraction (1\%) of UEs of the entire dataset.

$\bullet$ On top of the D-VAE, we propose a unified purification framework for countering various UEs. Extensive experiments demonstrate the remarkable performance of our method across CIFAR-10, CIFAR-100, and a 100-class ImageNet-subset, encompassing multiple poison types and different perturbation strengths, \textit{e.g.,} with only 4\% drop on ImageNet-subset compared to models trained on clean data.

\section{Related work}
\subsection{Data poisoning}
Data poisoning attacks~\citep{barreno2010security,goldblum2022dataset,yu2023backdoor}, involving the manipulation of training data to disrupt the performance of models during inference, can be broadly categorized into two main types: integrity attacks and availability attacks.
Integrity attacks aim to manipulate the model's output during inference~\citep{barreno2006can, xiao2015feature, zhao2017efficient}, \textit{i.e.,} backdoor attacks~\citep{gu2017badnets,schwarzschild2021just}, where the model behaves maliciously only when presented with data containing specific triggers. 
In contrast, availability attacks aim to degrade the overall performance on validation and test datasets~\citep{biggio2012poisoning,xiao2015feature}. 
%
{
Typically, such attacks inject poisoned data into the clean training set. Poisoned samples are usually generated by adding unbounded perturbations, and take only a fraction of the entire dataset~\citep{koh2017understanding, zhao2022clpa,10.5555/3618408.3619357}. These methods are primarily designed for malicious purposes, and the poisoned samples are relatively distinguishable.
}

{
\noindent \textbf{Unlearnable Examples.} Another recent emerging type is unlearnable examples (UEs)~\citep{dc,em}, where samples from the entire training dataset undergo subtle modifications (\textit{e.g.,} bounded perturbations $\Vert \boldsymbol{p} \Vert_{\infty} \le \frac{8}{255}$), and are correctly labeled. 
This type of attack, also known as perturbative availability poisoning attacks~\citep{iss}, can be viewed as a promising approach for data protection. Models trained on such datasets often approach random guessing performance on clean test data.
%
}
EM~\citep{em} employ error-minimizing noise {as perturbations}. NTGA~\citep{ntga} generate protective noise using an ensemble of neural networks modeled with neural tangent kernels. TAP~\citep{tap} employ targeted adversarial examples as UEs. REM~\citep{rem} focuses on conducting robust attacks against adversarial training. Subsequently, LSP~\citep{lsp} explore effecient and surrogate-free UEs, and extending the perturbations to be $\ell_2$ bounded. Recently, OPS~\citep{ops} introduce one-pixel shortcuts, which enhances the robustness to adversarial training and strong augmentations.

\subsection{Existing defenses}
\yuyi{
Defenses against UEs can be categorized into training-time defenses and pre-training purification, depending on interventions applied during or before the training phase. \citet{em} shown that UEs are robust to data augmentations, \textit{e.g.,} Mixup~\citep{zhang2018mixup}. \citet{tao2021better} find that adversarial training (AT) could mitigate poisoning effects, but it is computationally expensive and cannot {fully restore performance}. Building on the idea of AT, \citet{qin2023learning} employ adversarial augmentations (AA), but it still demands intensive training and does not generalize well to ImageNet-subset.
For pre-training defenses, ~\citet{iss} indicates that pre-filtering, \textit{e.g.,} gaussian smoothing, median filtering, show substantial effects but not comparable to AT. Instead, \citet{iss} propose image shortcut squeezing (ISS) including JPEG compression, grayscale, and bit depth reduction~\citep{bdr} to defense, while each technique does not fit all UEs approaches. Moreover, it is noted that low-quality JPEG compression, while effective for defense, significantly degrades image quality. AVATAR \citep{dolatabadi2023devil} and LEs \citep{jiang2023unlearnable} both employ a diffusion model for purification, but those methods require a substantial amount of additional clean data to train the diffusion model~\citep{ho2020denoising,song2020score}, making it impractical. 
LFU~\citep{lfu} is a hybrid method that adopt orthogonal projection to learn perturbations before training and employs strong augmentations during training due to incomplete purifications. However, it is constrained to UEs methods that adopt class-wise linear perturbations, limiting its applicability.
}

\begin{figure*}[t]
\begin{minipage}{1.0\linewidth}
\centerline{{\includegraphics[width=0.95\linewidth]{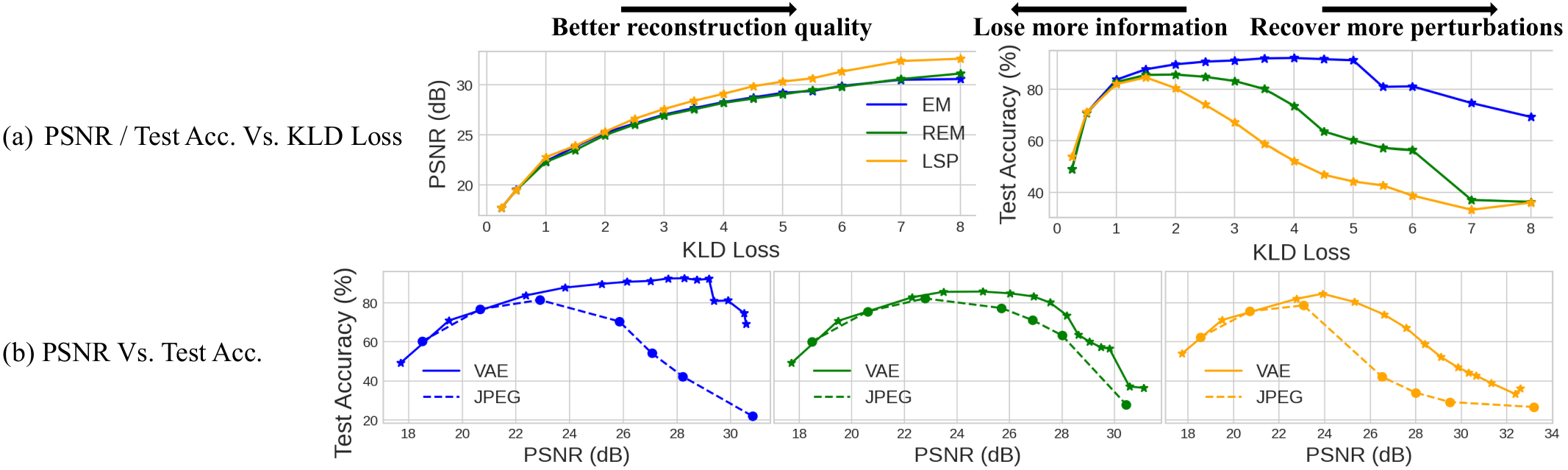}}}
\end{minipage}
\vspace{-4.5mm}
    \caption{
    (a): Results of VAEs: PSNR/Test Accuracy Vs. KLD Loss are assessed on the unlearnable CIFAR-10. 
    (b): Comparison between VAEs and JPEG compression: PSNR Vs. Test Accuracy.
    Note that we adopt JPEG with quality \{2,5,10,30,50,70,90\} to control the corruption levels. {We include EM, REM, and LSP as UEs methods.}}
    \label{trade-off}
\vspace{-2mm}
\end{figure*}

\section{Methodology}
\subsection{Preliminaries}
Formally, for UEs, all training data can be perturbed to some extent, while the labels should remain correct~\citep{dc,tap}. 
We introduce two parties: the attacker (also called the poisoner), and the victim. The attacker has the ability to perturb the victim's training data, \textit{i.e.,} from ($\boldsymbol{x_c}^{(i)}$, $y^{(i)}$) to ($\boldsymbol{x_c}^{(i)}+\boldsymbol{p}^{(i)}$, $y^{(i)}$).
The victim then trains a new model on the poisoned data, \textit{i.e.,} obtaining $\theta^{*}(\boldsymbol{p})$.
The attacker's success is determined by the accuracy of the victim model on clean data, \textit{i.e.,} maximizing the loss on clean data $\mathcal{L}({F}(\boldsymbol{x_c};\theta^{*}(\boldsymbol{p})),y)$.
The task to craft poisoning perturbations can be formalized into the following bi-level optimization problem:
\begin{equation}\small
\begin{split}
    & \max_{\boldsymbol{{p}} \in \mathcal{S}} \mathbb{E}_{({\boldsymbol{x_c}}, y) \sim \mathcal{D}}\big[\mathcal{L}({F}({\boldsymbol{x_c}};\theta^{*}(\boldsymbol{{p}})),y)\big], \quad \text{s.t.} ~\theta^{*}(\boldsymbol{{p}}) = \\
    & \argmin_{\theta} \sum\limits_{({\boldsymbol{x_c}^{(i)}}, y^{(i)}) \in \mathcal{T}}\mathcal{L}({F}({\boldsymbol{x_c}^{(i)}} + \boldsymbol{{p}}^{(i)};\theta),y^{(i)}),
    \label{availability_attack}
\end{split}
\end{equation}
where {${\boldsymbol{x_c}}$ is the clean data, and} $\mathcal{S}$ is the feasible region for perturbations, \textit{e.g.,} $\Vert\boldsymbol{p}\Vert_{\infty} \le \frac{8}{255}$.
By adding perturbations $\boldsymbol{{p}}^{(i)}$ to samples ${\boldsymbol{x_c}^{(i)}}$ from {the clean training dataset} $\mathcal{T}$ {to formulate the unlearnable training dataset $\mathcal{P}$}, the adversary aims to induce poor generalization of the trained model ${F}$ to the clean {test} dataset $\mathcal{D}$.

Conversely, data exploiters aim to obtain the learnable data by employing a mapping $\boldsymbol{g}$ such that:
\begin{equation}\small
\vspace{-0.5mm}
\begin{split}
     & \min_{\boldsymbol{g}} \mathbb{E}_{({\boldsymbol{x_c}}, y) \sim \mathcal{D}}\big[\mathcal{L}({F}({\boldsymbol{x_c}};\theta^{*}(\boldsymbol{g})),y)\big], \quad \text{s.t.} ~\theta^{*}(\boldsymbol{g}) = \\
     & \argmin_{\theta} \!\!\!\!\!\! \sum\limits_{{({\boldsymbol{x_c}^{(i)}} + \boldsymbol{{p}}^{(i)}, y^{(i)}) \in \mathcal{P}}} \!\!\!\!\!\!\! \mathcal{L}({F}(\boldsymbol{g}({\boldsymbol{x_c}^{(i)}} + \boldsymbol{{p}}^{(i)});\theta),y^{(i)}).
    \label{availability_attack_defense}
    \end{split}
\end{equation}
In this paper, we focus on pre-training purification, where $\boldsymbol{g}$ is applied for that purification, before training the classifier.

\noindent\textbf{Notations.} For better comprehension of the subsequent sections, we present notations for all remaining variables.
$\boldsymbol{\hat{p}}$ is the estimated perturbations by D-VAE. $\boldsymbol{\hat{x}}$/$\boldsymbol{z}$ is the reconstructed data, and the latents encoded by the VAE/D-VAE, respectively.
For the trainable modules, D-VAE consists of encoder $E_{\phi}$, decoder $D_{\theta_c}$, auxiliary decoder $D_{\theta_p}$, and class-wise embeddings ${\boldsymbol{u_y}}$.

\subsection{VAEs can effectively mitigate the impact of poisoning perturbations in UEs}\label{sec3.2}

The VAE maps the input to a lower-dimensional latent space, generating parameters for a variational distribution. 
The decoder reconstructs data from this latent space. 
The loss function combines a reconstruction loss (``distortion'') with a Kullback-Leibler (KL) divergence term (``rate''), acting as a limit on mutual information and serving as a compression regularizer~\citep{vae_rate}.

Since UEs have demonstrated vulnerability to compressions like JPEG, we first investigate whether a rate-constrained VAE can eliminate these perturbations and obtain the restored {learnable samples}. 
In essence, we introduce an updated loss function incorporating a rate constraint as follows (${\mathcal{P}}$ is the unlearnable dataset{, and $\boldsymbol{x}$ is the UEs}):
\vspace{-1.5mm}
\begin{equation}\small
\vspace{-2mm}
\begin{split}
    \!\!\! \mathcal{L}_{\text{VAE}} =  \!\!\! \sum_{\boldsymbol{x}, y \in {\mathcal{P}}} \! \underbrace{{\lVert \boldsymbol{x} - \boldsymbol{\hat{x}} \rVert}_2^2}_{\text{distortion}} + \lambda \underbrace{\text{max}(\text{KLD}(\boldsymbol{z},\mathcal{N}({\boldsymbol{0}},\boldsymbol{I})), \text{kld}_{\text{target}})}_{\text{rate constraint}},
    \label{rate_constrained loss}
\end{split}
\end{equation}
where the KLD Loss is formulated from~\citet{vae} and provided in the Appendix~\ref{kld_implementation}{, and the $\text{kld}_\text{target}$ serves as the target value for the KLD loss}.
We proceed to train the VAE on the unlearnable CIFAR-10. 
Subsequently, we report the accuracy on the clean test set achieved by a ResNet-18 trained on the reconstructed images. 
{
In Figure~\ref{trade-off}(a), reducing the KLD loss decreases reconstruction quality (measured by Peak Signal-to-Noise Ratio (PSNR) between $\boldsymbol{\hat{x}}$ and $\boldsymbol{{x}}$). This reduction can eliminate added perturbations and original valuable features. The right image of Figure~\ref{trade-off}(a) shows that increased removal of perturbations in $\boldsymbol{\hat{x}}$ correlates with improved test accuracy. However, heavily corrupting $\boldsymbol{\hat{x}}$ by further reducing $\text{kld}_\text{target}$ removes more valuable features, leading to a drop in test accuracy.
In Figure~\ref{trade-off}(b), the comparison with JPEG at various quality settings shows that when processed through VAEs and JPEG to achieve similar PSNR, test accuracy with VAEs is higher than JPEG. This suggests that VAEs are significantly more effective at eliminating perturbations than JPEG compression, when achieving similar levels of reconstruction quality.
}
%
%
Then, we delve into why VAEs can exhibit such preference.

\subsection{Theoretical Analysis and intrinsic characteristics}\label{vae_explaination}
Given that the feature extractor maps the input data to the latent space is pivotal for the classification conducted by DNNs, we conduct our analysis on the latent features ${\boldsymbol{v}}$.

\noindent \textbf{Hyperplane shift caused by attacks.} Consider the following binary classification problem with regards to {the features extracted from the data} ${\boldsymbol{v}} = ({\boldsymbol{v}}_c, {\boldsymbol{v}}_s^t)$ consisting of a predictive feature ${\boldsymbol{v}}_c$ of a Gaussian mixture ${\mathcal{G}}_c$ and a non-predictive feature ${\boldsymbol{v}}_s^t$ which follows:
\vspace{-1.5mm}
\begin{equation}\small
\begin{split}
    & y \stackrel{{u \cdot a\cdot r}}{\sim} \{0, 1\}, ~{\boldsymbol{v}}_c~{\sim}~\mathcal{N}({\boldsymbol{\mu}}_{c}^{y},{\boldsymbol{\Sigma}}_c), ~{\boldsymbol{v}}_s^{t}~{\sim}~\mathcal{N}({\boldsymbol{\mu}}^{t},{\boldsymbol{\Sigma}}^t), \\
    & \quad\quad\quad{\boldsymbol{v}}_c \indep {\boldsymbol{v}}_s^t, ~ \Pr(y=0)=\Pr(y=1).
\end{split}
    \label{distribution}
\end{equation}
\begin{proposition}
\vspace{-0.5mm}
For the {features} ${\boldsymbol{v}} = ({\boldsymbol{v}}_c, {\boldsymbol{v}}_s^t)$ following the distribution~(\ref{distribution}), the optimal separating hyperplane using a Bayes classifier is formulated by:
\vspace{-1mm}
\begin{equation}\small
    \boldsymbol{w}_c^{\top}({\boldsymbol{v}}_c^*-\frac{\boldsymbol{\mu}_c^0+\boldsymbol{\mu}_c^1}{2})=0, 
    ~\text{s.t.} ~\boldsymbol{w}_c = \boldsymbol{\Sigma}_c^{-1}(\boldsymbol{\mu}_c^0-\boldsymbol{\mu}_c^1).
\end{equation}
\label{proposition1}
\vspace{-5.5mm}
\end{proposition}
The proof is provided in Appendix~\ref{Proof_of_Proposition1}. 
Subsequently, we assume that a malicious attacker modifies ${\boldsymbol{v}}_s^t$ to ${\boldsymbol{v}}_s$ of the following distributions ${\mathcal{G}}_s$ to make it predictive for training a Bayes classifier:
\vspace{-1.5mm}
\begin{equation}\small
    y \stackrel{{u \cdot a\cdot r}}{\sim} \{0, 1\}, \quad {\boldsymbol{v}}_s~{\sim}~\mathcal{N}({\boldsymbol{\mu}}_{s}^{y},{\boldsymbol{\Sigma}}_s), \quad {\boldsymbol{v}}_c \indep {\boldsymbol{v}}_s.
    \label{poison_distribution}
\end{equation}

\begin{theorem}
Consider {features from the training data} for the Bayes classifier is modified from ${\boldsymbol{v}} = ({\boldsymbol{v}}_c, {\boldsymbol{v}}_s^t)$ in Eq.~\ref{distribution} to ${\boldsymbol{v}} = ({\boldsymbol{v}}_c, {\boldsymbol{v}}_s)$ in Eq.~\ref{poison_distribution}, the hyperplane is shifted with a distance given by:
\vspace{-1.5mm}
\begin{equation}\small
\begin{split}
   & \quad\quad d = {\lVert \boldsymbol{w}_s^{\top}({\boldsymbol{v}}_s-\frac{\boldsymbol{\mu}_s^0+\boldsymbol{\mu}_s^1}{2}) \rVert}_2\bigg/{{\lVert \boldsymbol{w}_c \rVert}_2}
    , \\
    & \text{s.t.} ~\boldsymbol{w}_c = \boldsymbol{\Sigma}_c^{-1}(\boldsymbol{\mu}_c^0-\boldsymbol{\mu}_c^1), ~\boldsymbol{w}_s = \boldsymbol{\Sigma}_s^{-1}(\boldsymbol{\mu}_s^0-\boldsymbol{\mu}_s^1).
\end{split}
    \label{hyperplane_shift_equation}
\end{equation}
\vspace{-4.5mm}
\label{theorem1}
\end{theorem}

The proof is provided in Appendix~\ref{Proof_of_Theorem1}. When conducting evaluations on the testing data that follows the same distribution as the clean data ${\boldsymbol{v}} = ({\boldsymbol{v}}_c, {\boldsymbol{v}}_s^t)$, with the term ${\boldsymbol{v}}_s$ in Eq.~\ref{hyperplane_shift_equation} replaced by ${\boldsymbol{v}}_s^t$, it leads to a greater prediction error if ${{\lVert \boldsymbol{w}_s \rVert}_2} \gg {{\lVert \boldsymbol{w}_c \rVert}_2}$. Theorem~\ref{theorem1} indicates that perturbations which create strong {attacks} tend to have a larger inter-class distance and a smaller intra-class variance.

\noindent \textbf{Error when aligning with a normal distribution.} Consider a variable ${\boldsymbol{v}} = ({v}_1, \dots, {v}_d)$ following a mixture of two Gaussian distributions ${\mathcal{G}}$:
\vspace{-1.5mm}
\begin{equation}\small
\begin{split}
    & \quad\quad\quad\quad\quad~ y \stackrel{{u \cdot a\cdot r}}{\sim} \{0, 1\}, ~{\boldsymbol{v}}~{\sim}~\mathcal{N}({\boldsymbol{\mu}}^{y},{\boldsymbol{\Sigma}}), \\
    & {v}_i~{\sim}~\mathcal{N}({\mu}_i^{y},{\sigma_i}), ~{v}_i \indep {v}_j, ~ \Pr(y=0)=\Pr(y=1), \\
  & \quad\quad\quad p_{{v}_i}({v}) = [\mathcal{N}({v};{\mu}_i^{0},{\sigma_i}) + \mathcal{N}({v};{\mu}_i^{1},{\sigma_i})]/2.
  \end{split}
\end{equation}
Each dimensional feature ${v}_i$ is also modeled as a Gaussian mixture. 
To start, we normalize each feature through a linear operation to achieve a distribution with zero mean and unit variance. 
The linear operation and the modified density function can be expressed as follows:
\vspace{-1.5mm}
\begin{equation}\small
\begin{split}
    &\quad z_i = \frac{{v}_i - \hat{\mu}_i}{\sqrt{{(\sigma_i)}^2+{(\delta_i)}^2}}, ~p_{z_i}({v}) = \frac{p_0({v})+p_1({v})}{2},\\
    &\quad\quad p_0({v}) = \mathcal{N}({v};-\hat{\delta}_i,\hat{\sigma}_i), ~ p_1({v}) = \mathcal{N}({v};\hat{\delta}_i,\hat{\sigma}_i),\\
    &\text{where} \quad\quad\quad \hat{\mu}_i =\frac{{\mu}^{0}_i+{\mu}^{1}_i}{2}, ~\delta_i =|\frac{{\mu}^{0}_i-{\mu}^{1}_i}{2}|,\\
    & \hat{\delta}_i=\delta_i/\sqrt{{(\sigma_i)}^2+{(\delta_i)}^2}, ~\hat{\sigma}_i=\sigma_i/\sqrt{{(\sigma_i)}^2+{(\delta_i)}^2}.
\end{split}
\label{normalized_distribution}
\end{equation}

\begin{theorem}
    Denote $r_i=\frac{\delta_i}{\sigma_i}>0$, the Kullback–Leibler divergence between $p_{z_i}({v})$ in (\ref{normalized_distribution}) and a standard normal distribution $\mathcal{N}({v};0,1)$ is bounded by:
\vspace{-1mm}
\begin{equation}\small
\vspace{-2mm}
\begin{split}
     \frac{\ln\!{(1+(r_i)^2)}}{2}\!-\!\ln\!2 \!\leq \!{\text{KLD}}(p_{z_i}({v})\Vert\mathcal{N}({v};0,1))
     \!\leq \!\frac{\ln\!{(1+(r_i)^2)}}{2},
\end{split}
\end{equation}
and observes the following property: 
\begin{equation}\small
\begin{split}
    \uparrow r_i & \quad \implies \quad \uparrow S(r_i)={\text{KLD}}(p_{z_i}({v})\Vert\mathcal{N}({v};0,1)).
\end{split}
\end{equation}
\label{theorem2}
\vspace{-5mm}
\end{theorem}
The proof for Theroem~\ref{theorem2} is provided in Appendix~\ref{Proof_of_Theorem2}.

\begin{algorithm*}[t]
    \caption{Two-stage purification framework of unlearnable examples with D-VAE}
    \scalebox{1.0}{
    \begin{minipage}{1\linewidth}
    \begin{algorithmic}
    \STATE \textbf{Input:} Unlearnable dataset ${\mathcal{P}^0}$ , D-VAE ($E_{\phi}$, $D_{\theta_c}$, $D_{\theta_p}$, ${\boldsymbol{u_y}}$), $\text{kld}_{\text{target}}$: $kld_1$, $kld_2$
    \STATE \textcolor{teal}{\# First stage: recover and remove heavy {perturbations} by training D-VAE with small $kld_1$}
    \STATE \text{Randomly initialize (${\phi}$, ${\theta_c}$, ${\theta_p}$, ${\boldsymbol{u_y}}$)}, and \text{using Adam to minimize Eq.~\ref{main_loss} on ${\mathcal{P}^0}$} with $kld_1$
    \STATE \text{Inference with trained VAE on ${\mathcal{P}^0}$, and save a new dataset ${\mathcal{P}^1}$} with sample $\boldsymbol{x}^1=\boldsymbol{x}^0 - \boldsymbol{\hat{p}}^0$
    \STATE \textcolor{teal}{\# Second stage: generate purified data by training D-VAE with larger $kld_2$}
    \STATE \text{Randomly initialize (${\phi}$, ${\theta_c}$, ${\theta_p}$, {$\boldsymbol{u_y}$)}}, and \text{using Adam to minimize Eq.~\ref{main_loss} on ${\mathcal{P}^1}$} with $kld_2$
    \STATE \text{Inference with trained VAE on ${\mathcal{P}^1}$, and save a new dataset ${\mathcal{P}^2}$} with sample $\boldsymbol{x}^2=\boldsymbol{x}^1 - \boldsymbol{\hat{p}}^1$
    \STATE \text{Inference with trained VAE on ${\mathcal{P}^2}$, and save a new dataset ${\mathcal{P}^3}$} with sample $\boldsymbol{x}^3 = \boldsymbol{\hat{x}}^2$
    \STATE \text{\textbf{Return} purified dataset ${\mathcal{P}^3}$}
    \end{algorithmic}
    \end{minipage}}
    \label{alg1}
\end{algorithm*}

\begin{remark}
%
The training of a VAE includes the process of mapping the data $\boldsymbol{x}$ to latents. We can break this process into two step2: 1) The encoder first map the $\boldsymbol{x}$ to lossless intermediate representations $z_i \sim p_{z_i}(v)$; 2) The encoder estimate (re-project/remap) the intermediate representations to a new distribution $\hat{P}$ subject to ${\text{KLD}}(\widehat{P}\Vert\mathcal{N}(0,1)) < \epsilon$.
According to Theorem 3.3, for each lossless intermediate representation $z_i$ with $r < S^{-1}(\epsilon)$, we can apply an identical mapping $\hat{P}=p_{z_i}({v})$ without requiring the step 2 mentioned above. In this way, the final representation $z_i$ is still lossless.
For the intermediate representation $z_i$ with $r_{v_i} > S^{-1}(\epsilon)$, we can see that through step 2, the final representation with $\hat{P}$ is forced to have a smaller $r$, and ideally to be almost equal to $S^{-1}(\epsilon)$. Basically, the error for the two distributions can be denoted as $\int_{-\infty}^{\infty}[\widehat{P}({v})-p_{z_i}({v})]^2d{v}$ in the distribution space. From this formulation, we can explicitly see that a larger gap $(r_{{v}_i} - S^{-1}(\epsilon))$ can also lead to larger estimation error. And the estimated $\widehat{P}$ is constrained to have a smaller $r$, making it less predictive for classification.
\label{remark1}
\end{remark}
Remark~\ref{remark1} indicates that perturbative patterns that make strong attacks tend to suffer from larger errors when estimating with distributions subject to the constraint on the KLD. Thus, the training of a rate-constrained VAE includes simulating the process of mapping the data to latent representations and aligning them with a normal distribution to a certain extent. 
The decoder learns to reconstruct the input data from the resampled latents $\boldsymbol{z}$. 
Consequently, the highly predictive shortcuts are subdued or eliminated in the reconstructed data $\boldsymbol{\hat{x}}$.

\begin{proposition}
    The conditional entropy of a Gaussian mixture ${\boldsymbol{v}}_s$ of~${\mathcal{G}}_s$ in Eq.~\ref{poison_distribution} is given by:
\begin{equation}\small
   H({\boldsymbol{v}}_s|y_i) = \frac{\text{dim}({\boldsymbol{v}}_s)}{2}(1+\ln(2\pi)) + \frac{1}{2}\ln|\boldsymbol{\Sigma}_s|,
\end{equation}
where $\text{dim}({\boldsymbol{v}}_s)$ is the dimensions of the features. If each feature ${{v}}_s^d$ is independent, then: 
\begin{equation}\small
     H({\boldsymbol{v}}_s|y_i) =\frac{\text{dim}({\boldsymbol{v}}_s)}{2}(1+\ln(2\pi)) + \sum_{d=1}^{\text{dim}({\boldsymbol{v}}_s)} \ln\sigma_s^d.
\end{equation}
\label{proposition2}
\vspace{-5mm}
\end{proposition}

As the inter-class distance $\Delta_s={{\lVert \boldsymbol{\mu}_s^0 - \boldsymbol{\mu}_s^1 \rVert}_2}$ is constrained to ensure the invisibility of the perturbations, {perturbations} in most UEs exhibit a relatively low intra-class variance. Proposition~\ref{proposition2} suggests that the class-conditional entropy of the perturbations is comparatively low. Adversarial poisoning~\citep{tap, sep} could be an exception since adversarial examples can maximize latent space shifts with minimal perturbation in the RGB space. However, the preference to be removed by VAE still holds.

\subsection{D-VAE: VAE with perturbations disentanglement}\label{sec_dvae}
\yuyi{
Given that the defender lacks groundtruth values for the perturbations $\boldsymbol{p}$, it is not possible to optimize $\boldsymbol{u_y}$ and $D_{\theta_p}$ to learn to predict $\boldsymbol{\hat{p}}$ directly by minimizing ${\lVert \boldsymbol{p} - \boldsymbol{\hat{p}} \rVert}_2^2$ during model training.
Expanding on the insights from Section~\ref{sec3.2} and Remark~\ref{remark1}, when imposing a low target value on the KLD loss, creating an information bottleneck on the latents $\boldsymbol{z}$, the reconstructed $\boldsymbol{\hat{x}}$ cannot achieve perfect reconstruction, making the added perturbations more challenging to be recovered in $\boldsymbol{\hat{x}}$. As a result, a significant portion of perturbations $\boldsymbol{p}$ persists in the residuals $\boldsymbol{x} - \boldsymbol{\hat{x}}$. 
}

\yuyi{
Following Proposition~\ref{proposition2}, the majority of perturbations associated with each class data exhibit relatively low entropy, suggesting that they can be largely reconstructed using representations with limited capacity. 
Considering that most perturbations are crafted to be sample-wise, we propose a learning approach that maps the summation of a trainable class-wise embedding $\boldsymbol{u_y}$ and the latents $\boldsymbol{z}$ to $\boldsymbol{\hat{p}}$ through an auxiliary decoder $D_{\theta_p}$.
To learn $\boldsymbol{u_y}$ and train $D_{\theta_p}$, we propose minimizing ${\lVert (\boldsymbol{x} - \boldsymbol{\hat{x}}) - \boldsymbol{\hat{p}} \rVert}_2^2$, as the residuals $\boldsymbol{x} - \boldsymbol{\hat{x}}$ contain the majority of the groundtruth $\boldsymbol{p}$ when imposing a low target value on the KLD loss.
}

The overall network is in Figure~\ref{dvae} (a), and the improved loss {to optimize the D-VAE ($E_{\phi}$, $D_{\theta_c}$, $D_{\theta_p}$, $\boldsymbol{u_y}$)} is given:
\begin{equation}\small
\begin{split}
    \mathcal{L}_{\text{D-VAE}} &=  \sum_{\boldsymbol{x, y} \in {\mathcal{P}}} \underbrace{{\lVert \boldsymbol{x} - \boldsymbol{\hat{x}} \rVert}_2^2}_{\text{distortion}} + \underbrace{{\lVert (\boldsymbol{x} - \boldsymbol{\hat{x}}) - \boldsymbol{\hat{p}} \rVert}_2^2}_{\text{recover {perturbations}}} \\
    & + \lambda \cdot \underbrace{\text{max}(\text{KLD}(\boldsymbol{z},\mathcal{N}({\boldsymbol{0}},\boldsymbol{I})), \text{kld}_{\text{target}})}_{\text{rate constraint}},
    \label{main_loss}
\end{split}
\end{equation}
{
$\text{where}~\boldsymbol{\mu}, \boldsymbol{\sigma} = E_{\phi}(\boldsymbol{x}),~\boldsymbol{z}~ \text{is sampled from}~\mathcal{N}(\boldsymbol{\mu},\boldsymbol{\sigma}),~\boldsymbol{\hat{x}} = D_{\theta_c}(\boldsymbol{z})$,$~\text{disentangled perturbations}~\boldsymbol{\hat{p}} = D_{\theta_p}(\boldsymbol{u_y}+\boldsymbol{z})$. 
%
}

\subsection{Purify UEs with D-VAE}\label{sec_framework}
Given the observations in Section~\ref{sec3.2} that a large KLD target fails to effectively surpress the {perturbations}, while a small one might significantly deteriorate the quality of reconstructed images, we introduce a two-stage purification framework as shown in Algorithm~\ref{alg1}.
In the first stage, we use a small $\text{kld}_\text{target}$ to train the VAE with the unlearnable dataset {\small{${\mathcal{P}^0}$}}. 
This approach allows us to reconstruct a significant portion of the perturbations. 
During inference, we subtract the input {\small{$\boldsymbol{x}^0$}} from {\small{${\mathcal{P}^0}$}} by the predicted perturbations {\small{$\boldsymbol{\hat{p}}^0$}} and obtain these modified images as {\small{${\mathcal{P}^{1}}$}}.

In the second stage, we set a larger $\text{kld}_\text{target}$ for training. After subtracting {\small{$\boldsymbol{x}^1$}} by {\small{$\boldsymbol{\hat{p}}^1$}} and saving it as {\small{$\boldsymbol{x}^2$}} in the first inference. 
Since the perturbations are learned in an unsupervised manner, it is challenging to achieve complete reconstruction. 
Hence, we proceed with a second inference and obtain the output {\small{$\boldsymbol{\hat{x}}^2$}} as the final result.

\noindent\textbf{Selection of $\boldsymbol{kld_1, kld_2}$.} Due to potential variations in the VAE's encoder and dataset's resolutions, we decide $\text{kld}_\text{target}$ empirically based on PSNR between the input $\boldsymbol{x}$ and the output $\boldsymbol{\hat{x}}$.
Specifically, though various UEs methods may utilize different norms to constrain the magnitude of perturbations, the disparity between the clean data $\boldsymbol{x_c}$ and poisoned data $\boldsymbol{x}$ in terms of PSNR is usually slightly above 30. Therefore, the selection of an appropriate $\text{kld}_\text{target}$ requires this prior information. Across multiple datasets, there is a basic strategy for selecting the proper $kld_1$ and $kld_2$ as follows:

\textbf{1)} As the first stage aims to remove the majority of perturbations, we need to adopt a small $kld_1$ to ensure that PSNR between $\boldsymbol{x}$ and $\boldsymbol{\hat{x}}$ is low. Thus, most perturbations are preserved in $\boldsymbol{x} - \boldsymbol{\hat{x}}$, thereby leading to a better estimation of $\boldsymbol{\hat{p}}$. We choose PSNR value to fall between 20 and 22, and experiment with some selections of $kld_1$ to achieve this.

\textbf{2)} For the second stage, where our goal is to obtain poison-free data while maintaining high quality, it is ideal for the PSNR between $\boldsymbol{x}$ and $\boldsymbol{\hat{x}}$ to range between 28 and 30. This range is slightly below the typical PSNR value between $\boldsymbol{x_c}$ and $\boldsymbol{x}$ (usually around 30). Then we experiment with various selections of $kld_2$ to achieve this.

\textbf{3)} Essentially, the selection process outlined above is dependent on both the dataset and the encoder structure of the VAE. Consequently, for the same dataset but with different UEs methods, we opt for the same $kld_1$ and $kld_2$.

\section{Experiments}
\subsection{Experimental Setup}
\noindent \textbf{Datasets and models.} 
We choose three commonly used datasets: CIFAR-10, CIFAR-100~\citep{cifar}, and a subset of ImageNet~\citep{imagenet} with the first 100 classes. 
For CIFAR-10 and CIFAR-100, we maintain the original size of $32 \times 32$. 
Regarding the ImageNet subset, we follow prior research~\citep{em}, and resize the image to $224 \times 224$.
In our main experiments, we adopt the ResNet-18~\citep{resnet} model as both the surrogate and target model. 
To evaluate transferability, we include various classifiers, such as ResNet-50, DenseNet-121~\citep{densenet}, MobileNet-V2~\citep{mobilenetv2}. 

\noindent \textbf{Unlearnable examples.} 
We examine several representative UEs methods with various perturbation bounds. 
The majority of methods rely on a surrogate model, including NTGA~\citep{ntga}, EM~\citep{em}, REM~\citep{rem}, TAP~\citep{tap}, SEP~\citep{sep}, and employ the $\ell_{\infty}$ bound. 
On the other hand, surrogate-free methods such as LSP~\citep{lsp} and AR~\citep{ar} utilize the $\ell_2$ bound.
Additionally, OPS~\citep{ops} utilizes the $\ell_0$ bound. The diversity of these attacking methods can validate the generalization capacity of our proposed purification framework.

\noindent \textbf{Competing defensive methods.} 
We include two training-time defenses: adversarial training (AT) with $\epsilon = 8/255$~\citep{wen2022adversarial} and adversarial augmentations (AA)~\citep{qin2023learning}. Among the pre-training methods, we include ISS~\citep{iss}, consisting of bit depth reduction (BDR), Grayscale, and JPEG, as well as AVATAR~\citep{dolatabadi2023devil} (denoted as AVA.), which employs a diffusion model trained on the clean CIFAR-10 dataset to purify. We also include LFU~\citep{lfu}, a hybrid defense that utilizes orthogonal projection to learn perturbations. It also requires strong augmentations during training due to incomplete purifications. For a fair comparison, we choose to report the test accuracy from the last epoch. More details are in the Appendix~\ref{competing_defenses}.

\noindent \textbf{Model Training.} 
To ensure consistent training procedures for the classifier, we have formalized the standard training approach. 
For CIFAR-10, we use 60 epochs, while for CIFAR-100 and the ImageNet, 100 epochs are allowed. 
In all experiments, we use SGD optimizer with an initial learning rate of 0.1 and the CosineAnnealingLR scheduler, keeping a consistent batch size of 128.
For D-VAE training on unlearnable CIFAR-10, we use a KLD target of 1.0 in the first stage and 3.0 in the second stage, with only a single $\times 0.5$ downsampling to preserve image quality. 
For the CIFAR-100, we maintain the same hyperparameters as CIFAR-10, except for setting $kld_2$ to 4.5.
For ImageNet-subset, which has higher-resolution images, we employ more substantial downsampling ($\times 0.125$) in the first stage and set a KLD target of 1.5, while the second stage remains the same as with CIFAR. 
When comparing the unlearnable input and the reconstructed output, these hyperparameters yield PSNRs of around 28 for CIFAR and 30 for ImageNet.
In Appendix~\ref{Selection_of_various_kld}, we showcase that our method is tolerant to the selection of $kld_1$ and $kld_2$.

\vspace{1mm}
\subsection{Validate the effectiveness of the disentanglement}\label{sec4.2}
UEs can be analyzed from the standpoint of shortcuts~\citep{lsp}. 
It has been empirically shown that models trained on the unlearnable training data have a tendency to memorize the perturbations, and attain high accuracy when testing on data that has same perturbations~\citep{iss}.

In this section, we aim to illustrate that the disentangled perturbations remain effective as potent {attacks} and can be regarded as equivalent to the original unlearnable dataset ${{\mathcal{P}}}$. 
{
Initially, we look into the amplitude of the perturbations in terms of $\ell_2$-norm.
From Table~\ref{dis_norm} in the Appendix~\ref{disentangled_poison_patterns}, the amplitude of groundtruth $\boldsymbol{p}$ is around 1.0 for LSP and AR, and about 1.5 for others. The generated $\boldsymbol{\hat{p}}$ has an amplitude of about 1.8 for OPS and around 0.7 to 1.0 for others. Notably, the amplitude of $\boldsymbol{\hat{p}}$ is comparable to that of $\boldsymbol{{p}}$, with $\boldsymbol{\hat{p}}$ being slightly smaller than $\boldsymbol{{p}}$ except for OPS. 
\yuyi{
The visual results of the normalized perturbations can be seen in Figure~\ref{dvae}, and we observe the visual similarity between $\boldsymbol{\hat{p}}$ and $\boldsymbol{{p}}$, especially for LSP and OPS. More details and visual results are in the Appendix~\ref{disentangled_poison_patterns} and Appendix~\ref{more_visual_results}, respectively.
}
}

Subsequently, we construct a new unlearnable dataset denoted as {{${{\widehat{\mathcal{P}}}}$}} by incorporating the disentangled perturbations ${\boldsymbol{\hat{p}}}$ into the clean training set {{${{\mathcal{T}}}$}}. We proceed to train a model using {{${{\widehat{\mathcal{P}}}}$}}, and subsequently evaluate its performance on three distinct sets: the clean training set {{${{\mathcal{T}}}$}}, the clean testing set {{${{\mathcal{D}}}$}}, and the original unlearnable dataset {{${{\mathcal{P}}}$}}. 
From the results in Table~\ref{validate}, it becomes apparent that the reconstructed dataset continues to significantly degrade the accuracy on clean data. 
{
In fact, compared to the attacking performance of {{${{{\mathcal{P}}}}$}} in Table~\ref{cifar10} and Table~\ref{cifar100}, {{${{\widehat{\mathcal{P}}}}$}} even manages to achieve an even superior attacking performance in most instances with less amplitude.
}
During testing on the original unlearnable dataset, the accuracy levels are notably high, often approaching 100\%. This outcome serves as an indicator of the effectiveness of the disentanglement process.

\begin{table}[t]\footnotesize
    \centering
\setlength\tabcolsep{4.5pt}
    \caption{Testing accuracy (\%) of models trained on reconstructed unlearnable dataset {{${{\widehat{\mathcal{P}}}}$}}.  }
    \centering
    \scalebox{0.9}{
    \begin{tabular}{ c | c | c  c  c  c  c  c }
    \toprule
     \multirow{1}*{Datasets} & Test Set& EM & REM & NTGA & LSP & AR & OPS \\
    \midrule
     \multirow{3}*{CIFAR-10} &${\mathcal{T}}$&9.7&19.8& 29.2&15.1 & 13.09 & 18.5\\
     &${\mathcal{D}}$&9.6
&19.5&28.6&15.3 & 12.9 & 18.7\\
     &${\mathcal{P}}$&91.3
&99.9&99.9&99.9 & 100.0 & 99.7\\
     \midrule
     \multirow{3}*{CIFAR-100}&${\mathcal{T}}$&1.4&6.4&- &4.2 & 1.6 & 11.2 \\
     & ${\mathcal{D}}$&1.3
&7.6& -&4.0 & 1.6 & 10.7\\
    & ${\mathcal{P}}$&98.8
    &96.4&- & 99.1 & 100.0 & 99.5\\
    \bottomrule
    \end{tabular}
    \label{validate}}
\vspace{-3mm}
\end{table}

\begin{table*}[t]\footnotesize
    \caption{
    Clean test accuracy (\%) of models trained on the unlearnable CIFAR-10 dataset and with our proposed method Vs. other defenses. 
    Our results on additional classifiers are at the rightmost. RN, DN, and MN denote ResNet, DenseNet, and MobileNet, respectively.
    }
    \setlength\tabcolsep{5.0pt}
    \centering
    \scalebox{0.85}{
    \begin{tabular}{ c | c | c | c c | c  c  c  c c | c || c c c}
    \toprule
     \multirow{2}*{Norm} & {UEs} / Countermeasures & w/o & AT & AA & BDR & Gray &JPEG & AVA. &LFU& Ours & RN-50 & DN-121 & MN-v2 \\
     \cmidrule{2-14}
       & Clean (no poison) & \textbf{94.57} & 85.17 & {92.27}& 88.95 & 92.74 & 85.47& 89.61 & 86.78 & 93.29 & 93.08 & 93.73 & 83.61\\
    \midrule
     \multirow{5}*{$\ell_{\infty} = \frac{8}{255}$} &NTGA~\citep{ntga}&11.10&83.63& 77.92&57.80 & 65.26 & 78.97 & 80.72&82.21 & \textbf{89.21} &88.96& 89.28&78.72\\
     &EM~\citep{em}&12.26
&84.43& 67.11& 81.91 & 19.50 & 85.61 & 89.54& 65.17& \textbf{91.42} & 91.62 & 91.64 & 81.10\\
     &TAP~\citep{tap}&25.44
&83.89 & 55.84&80.18 & 21.50 & 84.99 & 89.13& 53.46& \textbf{90.48} &90.50&90.51&81.28\\
     &REM~\citep{rem}&22.43
&86.01& 64.99& 32.36 & 62.35 & 84.40 & 86.06& 33.81 & \textbf{86.38} &85.91&86.74&79.27\\
     &SEP~\citep{sep}&6.63
&83.48& 61.07& 81.21 & 8.47 & 84.97 & 89.56&  74.14&\textbf{90.74} &90.86&90.76&80.98\\
     \midrule
     \multirow{2}*{$\ell_{2} = 1.0$}& LSP~\citep{lsp}&13.14&84.56& 80.39& 40.25 & 73.63 & 79.91 & 81.15& 87.76&\textbf{91.20} & 90.15&91.10&80.26\\
     & AR~\citep{ar}&12.50
&82.01& 49.14&29.14 & 36.18 & 84.97 & 89.64& 23.51& \textbf{91.77} &90.53&90.99&82.26\\
     \midrule
     {$\ell_{0} = 1$}& OPS~\citep{ops}&22.03
& 9.48 & 64.02& 19.58 & 19.43 & 77.33 & 71.62&86.46 &\textbf{88.95} & 88.10 & 88.78 & 81.40\\
\midrule
\multicolumn{2}{c|}{Mean (except clean)}&15.69
& 74.68 & 65.06& 52.80 & 38.29 & 82.64 & 84.67& 63.19&\textbf{90.01} & 89.58 & 89.98 & 80.66\\
    \bottomrule
    \end{tabular}}
    \label{cifar10}
    \vspace{-1.5mm}
\end{table*}

\begin{table*}[t]\footnotesize
\vspace{-1.2mm}
\parbox{.36\linewidth}{
  \centering
  \caption{Performance on CIFAR-100.  }
  \scalebox{0.85}{
  \setlength\tabcolsep{1.9pt}
    \centering
    \begin{tabular}{ c | c | c c | c  c c| c}
    \toprule
     {UEs} & w/o & AT & AA & ISS& AVA. & LFU & Ours \\
     \midrule
       Clean & \textbf{77.61} & 59.65 & 69.09& 71.59 & 61.09 & 33.12& 70.72\\
    \midrule
     EM&12.30
&59.07& 42.89& 61.91 & 61.09& 29.54&\textbf{68.79}\\
     TAP&13.44
&57.91& 35.10& 57.33& 60.47& 29.90&\textbf{65.54}\\
     REM&16.80
&59.34& 50.12& 58.13 & 60.90& 31.06&\textbf{68.52}\\
     SEP&4.66
&57.93& 27.77& 57.76 & 59.80& 32.03&\textbf{64.02}\\
     LSP&2.91&58.93& 53.28& 53.06 & 52.17 & 34.61&\textbf{67.73}\\
     AR&2.71
&58.77& 26.77& 56.60 & 60.33& 30.09&\textbf{63.73}\\
     OPS&12.56
& 7.28 & 36.78& 54.45 & 44.24& 30.40&\textbf{65.10}\\
\midrule
     Mean & 9.34 & 51.32&38.96&57.03&57.00& 31.09&\textbf{66.20}\\
    \bottomrule
    \end{tabular}
    \label{cifar100}
}}
\parbox{.27\linewidth}{
  \centering
  \caption{Performance on 100-class ImageNet-subset.  }
    \centering
     \scalebox{0.85}{
     \setlength\tabcolsep{1.9pt}
    \begin{tabular}{c | c | c c | c| c}
    \toprule
     {UEs} & w/o & AT & AA & ISS & Ours\\
     \cmidrule{1-6}
       Clean & \textbf{80.52} & 55.94 & 71.56& 76.92 & 76.78\\
    \midrule
     EM&1.08
&56.74& 3.82& 72.44 &\textbf{74.80}\\
     TAP&12.56
&55.36&71.38&73.24 &\textbf{76.56}\\
     REM&2.54
&59.34&20.92&58.13 &\textbf{72.56}\\
     LSP&2.50&58.93&46.58&53.06 & \textbf{76.06}\\
     \midrule
     Mean&4.67&57.59&35.68&64.21&\textbf{75.00}\\
    \bottomrule
    \end{tabular}
    \label{imagenet-mini}
}}
\hspace{0.5mm}
\parbox{0.35\linewidth}{
    \centering
\setlength\tabcolsep{1.9pt}
    \caption{Performance on unlearnable CIFAR-10 with larger bounds: {$\ell_{\infty} = \frac{16}{255}$} and {$\ell_{2} = 4.0$}.  }
    \centering
    \scalebox{0.85}{
\begin{tabular}{ c | c | c c | c c c| c}
    \toprule
     {UEs} & w/o & AT & AA & ISS & AVA. & LFU & Ours\\
    \midrule
     EM&10.09
&84.02&49.23&83.62 &85.61& 78.78& \textbf{91.06}\\
     TAP&18.45
&83.46&52.92&84.98 & 89.43& 22.23&\textbf{90.55}\\
     REM&23.22
&35.41&50.92&75.50 & 52.26& 83.10&\textbf{79.18}\\
     SEP&12.05
&83.98&56.71&85.00 & 88.96& 70.49&\textbf{90.93}\\
     LSP&15.45&79.10&59.10&41.41 &41.70 & 44.48&\textbf{86.43}\\
     \midrule
     Mean&15.85&73.19&53.77&74.10&71.59& 59.81&\textbf{87.63}\\
    \bottomrule
    \end{tabular}
    \label{larger_perturbation}}}
\vspace{-1mm}
\end{table*}

\subsection{Experimental results on UEs purification}
\noindent \textbf{CIFAR-10 UEs purification.} 
To evaluate the effectiveness of our purification framework, we conducted initial experiments on CIFAR-10. 
As shown in Table~\ref{cifar10}, our method consistently provides comprehensive protection against UEs with varying perturbation bounds and attack methods. 
In contrast, ISS relies on multiple simple compression techniques and requires adaptive selection of these methods, resulting in subpar defense performance. 
Notably, when compared to adversarial training, our method achieved an approximately 6\% improvement in performance. 
Even compared with AVATAR, which utilizes a diffusion model trained on the clean CIFAR-10 data, our methods achieve superior performance across all attack methods.
Our methods excel, especially on OPS attacks, which often perturb a pixel to its maximum value, creating a robust shortcut that evades most defenses. Our approach can effectively disentangle the majority of these additive perturbations in the first stage. The subsequent subtraction process can significantly mitigate the {attacks}, resulting in the poison-free data in the second stage.
The performance of training different classification models on our purified data is reported in the rightmost column of Table~\ref{cifar10}. As can be observed, our method indeed restores the learnability of data samples.

\noindent \textbf{CIFAR-100/ImageNet-subset UEs purification.} 
We then expand our experiments to include CIFAR-100 and a 100-class ImageNet subset. 
Due to the resource-intensive nature of the experiments, we focused on four representative attack methods for the ImageNet subset. Note that for ISS, we report the best accuracy among three compressions. 
The results, as presented in Table~\ref{cifar100} and Table~\ref{imagenet-mini}, re-confirm the overall effectiveness of our purification framework.

\noindent \textbf{Experiments on larger perturbations.} In our additional experiments, we introduced UEs with larger perturbation bounds. The outcomes on CIFAR-10 are outlined in Table~\ref{larger_perturbation}. 
It is worth noting that our method exhibits a high degree of consistency, with almost no performance degradation on EM, TAP, and SEP, and only a slight decrease on REM and LSP. However, it proves to be a challenging scenario for the competing methods to effectively address.

\yuyi{
\noindent \textbf{Comparison of existing defenses.} We offer a comparison of existing defenses and our approach. As shown in Table~\ref{summary_existing_defenses}, our method belongs to the pre-training purification, and requires no external clean data. It consistently outperforms across all UEs and datasets. Furthermore, among all competing defenses, LFU is the only one capable of learning and disentangling class-wise linear perturbations, applicable to LSP and OPS. However, due to incomplete purification, LFU also adopts strong augmentations and CutMix during training. In contrast, our method effectively disentangles perturbations in almost all UEs, except for adversarial poisoning including TAP and SEP, detailed in Proposition~\ref{proposition2} and Section~\ref{sec4.2}. More comparison is in the Appendix~\ref{non-variational},\ref{computation_cost}.
}

\begin{table}[t]\footnotesize
\vspace{-3mm}
    \centering
\setlength\tabcolsep{1.7pt}
    \caption{Comparison of existing defenses and our method. Performance drop is on CIFAR-10 dataset compared to clean one. }
    \centering
    \scalebox{0.85}{
    \begin{tabular}{c | c c c c c c c}
    \toprule
    {Characteristics} & AT & AA & ISS & AVA. & LFU & Ours \\
    \midrule
    {\shortstack{Pre-training purification}} & \XSolidBrush & \XSolidBrush & \Checkmark & \Checkmark & \Checkmark & \Checkmark \\
    {\shortstack{Training-phase interventions}} & \Checkmark & \Checkmark & \XSolidBrush & \XSolidBrush & \Checkmark & \XSolidBrush \\
    {\shortstack{No external clean data}} & \Checkmark & \Checkmark & \Checkmark & \XSolidBrush & \Checkmark & \Checkmark \\
    {\shortstack{Consistence on various UEs}} & \XSolidBrush & \XSolidBrush & \XSolidBrush & \Checkmark & \XSolidBrush & \Checkmark \\
    {\shortstack{UEs types that can be disentangled}} & 0/8 & 0/8 & 0/8 & 0/8 & 2/8 & \textbf{6}/8 \\
    {Mean performance drop (\%) $\downarrow$} & 19.89 & 29.51 & 11.93 & 9.90 &  31.38& \textbf{4.56} \\
    \bottomrule
\end{tabular}}
    \label{summary_existing_defenses}
\vspace{-4.0mm}
\end{table}

\begin{table}[t]\footnotesize
    \setlength\tabcolsep{1.6pt}
    \caption{Ablation study on the two-stage purification framework. s1/s2 denote the 1st and 2nd stage. i1/i2/i3 denote the 1st, 2nd and 3th inference. \ding{176} is a method where, after s1, we execute an operation same to i3, employing the D-VAE trained in s1.}
    \centering
    \scalebox{0.85}{
    \begin{tabular}{ c | c  c  c  c  c  c  c c | c}
    \toprule
     Method & NTGA & EM & TAP & REM & SEP & LSP & AR & OPS & Mean\\
    \midrule
     \ding{172}w/o s1 
     &78.62
&\textbf{91.85}&\textbf{90.97} & 82.06 & 90.76& 66.76 & 91.39&51.71 & 80.52\\
     \ding{173}w/o i2 in s2 &87.44&91.18
&90.70&85.21 & \textbf{90.79} & 90.63 & 91.31 & 84.92 & 89.02\\
\ding{174}w/o s2 &12.78&78.96&21.12&25.44	&4.83&93.47&11.49&41.57&36.21\\
\ding{175}w/o i3&13.87&80.77&23.02&23.84&5.29&93.58&14.23&66.39&40.12\\
\ding{176}&80.98&83.37&84.14&83.48&83.32&83.91&84.22&84.06&83.44\\
     Ours &\textbf{89.21}&91.42
&90.48&\textbf{86.38} & 90.74 & \textbf{91.20} & \textbf{91.77} & \textbf{88.95} & \textbf{90.01}\\
    \bottomrule
    \end{tabular}}
    \label{ablation}
\vspace{-1mm}
\end{table}

\subsection{Ablation Study}
\vspace{-1mm}
In this section, we conduct an ablation study on our two-stage purification framework. 
As shown in Table~\ref{ablation}, \ding{172} shows that the subtraction process in s1 plays a critical role in mitigating certain {attacks}, including NTGA, LSP, and OPS. 
\yuyi{\ding{173} shows that the subtraction (i2) in s2 can further improve the performance.}
This is particularly evident for LSP, which introduces smooth colorized blocks, and OPS, which perturbs a single pixel to a maximum value, making them challenging to remove when subjected to a moderate KLD target.
\ding{174} indicate that the reduction operation in s1 can partially mitigate the effects of poisoning perturbations, particularly demonstrating effectiveness against LSP. 
Since LSP and OPS adopt class-wise perturbations, the estimation of $\boldsymbol{\hat{p}}$ is more accurate and complete. 
%
\ding{175} reveal that an additional reduction operation in s2 leads to further elimination of perturbations and enhances performance compared to \ding{174}.
In contrast to \ding{174}, \ding{176} indicate that the output of D-VAE contains fewer poisoning perturbations. However, this reduction in perturbations comes at a cost: the small $\text{kld}_\text{target}$ in s1 results in outputs with lower reconstruction quality and a greater loss of useful information. In contrast, the outcomes of experiment \ding{173} are superior, as it adopts a larger $\text{kld}_\text{target}$.

\begin{table}[t]\footnotesize
    \centering
  \caption{Performance of detecting UEs or increasing UEs with various poisoning ratios on CIFAR-10 dataset.  }
  \scalebox{0.85}{
  \setlength\tabcolsep{3.5pt}
    \centering
 \begin{tabular}{ c | c | c  c  c  c || c | c }
    \toprule
     UEs& \multicolumn{5}{c||}{Detecting UEs} & \multicolumn{2}{c}{Increasing UEs}\\
    \midrule
     {Attacks} & \multirow{1}*{Ratio} &  Acc. & Recall & Precision & F1-score & Ratio & Test Acc.\\
    \midrule
     EM&\multirow{2}*{0.2} &0.918&1.0&0.709 & 0.830&\multirow{2}*{0.01} & 0.1009 \\
     LSP&&0.777
&1.0&0.472 & 0.641& & 0.1558\\
     \midrule
     EM&\multirow{2}*{0.4} &0.939&1.0&0.869 & 0.930&\multirow{2}*{0.02}& 0.1011\\
     LSP&&0.905
&1.0&0.807 & 0.893& & 0.1633\\
\midrule
     EM&\multirow{2}*{0.6} &0.961&1.0&0.938 & 0.968&\multirow{2}*{0.04} & 0.1229\\
     LSP&&0.941
&0.999&0.912 &  0.954& & 0.1405 \\
\midrule
EM&\multirow{2}*{0.8} &0.982&1.0&0.978 & 0.989&\multirow{2}*{0.08} & 0.1001\\
     LSP&&0.973
&1.0&0.968 & 0.984&& 0.1763\\
    \bottomrule
    \end{tabular}
    \label{detection_poison_generation}}
    \vspace{-3mm}
\end{table}

\section{Discussion}
\vspace{-1mm}
\subsection{Partial poisoning and UEs detection}
\vspace{-1mm}
In practical scenarios, it is often the case that only a fraction of the training data can be contaminated. 
Therefore, in line with previous research~\cite{iss}, we evaluate these partial poisoning scenarios by introducing UEs to a specific portion of the training data and subsequently combining it with the remaining clean data for training the target model. 
We conduct experiments on CIFAR-10 dataset.

When examining the first stage as outlined in Algorithm~\ref{alg1}, we observe that even the perturbations learned for the clean samples can potentially serve as poisoning attacks. This could be caused by the constrained representation capacity of the class-wise embedding.
In essence, building upon this discovery, we have the capability to create a new unlearnable dataset denoted as {{${{{\widehat{\mathcal{P}}}^0}}$}}, where each sample is formed as {{$\boldsymbol{x}^0+\boldsymbol{\hat{p}}^0$}}. 
Models trained on {{${{{\widehat{\mathcal{P}}}^0}}$}} tend to achieve high prediction accuracy on the unlearnable samples but perform notably worse on the clean ones.
Consequently, we can employ this metric as a means to detect the presence of unlearnable data, and the detection performance is outlined in Table~\ref{detection_poison_generation}. Notably, our detection method attains high accuracy, with an almost 100\% recall rate.
Subsequently, to address the issue of partial poisoning in datasets, we can adopt a detection-purification approach. 
The performances of models trained on the purified data are presented in Table~\ref{partial}.

\subsection{Increasing the amounts of UEs}
\vspace{-1mm}
In this section, we investigate whether our proposed disentanglement approach can help increase the amount of UEs once the attacker acquires additional clean data.
We conduct experiments on CIFAR-10 dataset by generating UEs, denoted as {{${\mathcal{P}}_{(0)}$}}, using a small ratio of the dataset, while leaving the remaining clean data {{$\mathcal{T}_{(1)}$}} untouched.
Subsequently, after training the D-VAE on {{${\mathcal{P}}_{(0)}$}}, we conduct inference on the {{$\mathcal{T}_{(1)}$}}. 
The addition of {{$\boldsymbol{\hat{p}}_{(1)}$}} to the clean data in {{$\mathcal{T}_{(1)}$}} results in a unlearnable dataset {{${\mathcal{P}}_{(1)}$}}. 
By combining {{${\mathcal{P}}_{(0)}$}} and {{${\mathcal{P}}_{(1)}$}} to create {{${\mathcal{P}}$}}, we proceed to train a classifier. The accuracy on the clean test set ${\mathcal{D}}$ are reported in Table~\ref{detection_poison_generation}. 
It is evident that training D-VAE with just 1\% UEs is adequate for generating additional UEs. More results are in the Appendix~\ref{more_partial_poisoning}.

\begin{table}[t]\footnotesize
\caption{Clean testing accuracy (\%) of models trained on the unlearnable CIFAR-10 dataset with different poisoning ratios.  }
\centering
\scalebox{0.85}{
\setlength\tabcolsep{4.2pt}
\begin{tabular}{ c | c | c  c  c  c  c c c}
\toprule
 {Ratio} & Counter &EM&TAP&REM&SEP&LSP&AR&OPS \\
\midrule
\multirow{2}*{0.2}& JPEG
& 85.03& 85.1 & 84.64& 85.34& 85.22 & 85.31 & 85.12\\
& Ours& \textbf{93.50}
&\textbf{90.55}&\textbf{92.24} & \textbf{90.86}& \textbf{93.20}&\textbf{92.77}&\textbf{93.15}\\
\midrule
 \multirow{2}*{0.4}& JPEG&85.31
&85.60&84.90& 85.22&85.34&85.29&84.89\\
& Ours&\textbf{93.03}
&\textbf{90.78}&\textbf{92.51} & \textbf{90.63} & \textbf{92.85} & \textbf{91.83} & \textbf{93.29}\\
\midrule
\multirow{2}*{0.6}& JPEG&85.40
&84.92& 84.62& 85.06 & 84.26 & 85.33 & 84.43\\
& Ours&\textbf{93.02}
&\textbf{90.93}&\textbf{92.23} & \textbf{91.04} & \textbf{92.16} & \textbf{91.41} & \textbf{92.13}\\
\midrule
\multirow{2}*{0.8}& JPEG&85.31
&85.34&84.97 & 85.06 & 83.02 & 84.87 & 83.01\\
& Ours&\textbf{92.26}
&\textbf{91.10}&\textbf{90.86} & \textbf{91.79} & \textbf{92.16} & \textbf{91.70} & \textbf{92.16}\\
\bottomrule
\end{tabular}
\label{partial}}
\vspace{-3mm}
\end{table}

\section{Conclusion}
\vspace{-1mm}
In this paper, we initially demonstrate that rate-constrained VAEs exihibit a natural preference for removing {poisoning perturbations} in unlearnable examples (UEs) by constraining the KL divergence in the latent space. 
We further provide a theoretical explanation for this behavior. 
Additionally, our investigations reveal that {perturbations} in most UEs have a lower class-conditional entropy, and can be disentangled by learnable class-wise embeddings and an auxiliary decoder. 
Building on these insights, we introduce the D-VAE, capable of disentangling the perturbations, and propose a two-stage purification framework that offers a consistent defense against UEs. 
Extensive experiments show the remarkable performance of our method across CIFAR-10, CIFAR-100, and ImageNet-subset, with various UEs and varying perturbation levels, \textit{i.e.,} only around 4\% drop on ImageNet-subset compared to models trained on clean data.
We plan to extend our work to purify UEs that target unsupervised learning scenarios in our future work.

\noindent\textbf{Acknowledgement.} 
This work was done at Rapid-Rich Object Search (ROSE) Lab, School of Electrical \& Electronic Engineering, Nanyang Technological University. This research is supported in part by the NTU-PKU Joint Research Institute and the DSO National Laboratories, Singapore, under the project agreement No. DSOCL22332.
\clearpage
\newpage
\section*{{Impact Statement}}
{
In summary, our paper presents a effective defense strategy against unlearnable examples (UEs), which aim to undermine the overall performance on validation and test datasets by introducing imperceptible perturbations to training examples with accurate labels.
UEs are viewed as a promising avenue for data protection, particularly to thwart unauthorized use of data that may contain proprietary or sensitive information. However, these protective methods pose challenges to data exploiters who may interpret them as potential threats to a company's commercial interests. 
Consequently, our method can be employed for both positive usage, such as neutralizing malicious data within a training set, and negative purpose, including thwarting attempts at preserving data privacy.
Our proposed method not only serves as a powerful defense mechanism but also holds the potential to be a benchmark for evaluating existing attack methods. We believe that our paper contributes to raising awareness about the vulnerability of current data protection techniques employing UEs. 
This, in turn, should stimulate further research towards developing more reliable and trustworthy data protection techniques.
}

\nocite{langley00}

\bibliography{example_paper}

\begin{thebibliography}{49}
\providecommand{\natexlab}[1]{#1}
\providecommand{\url}[1]{\texttt{#1}}
\expandafter\ifx\csname urlstyle\endcsname\relax
  \providecommand{\doi}[1]{doi: #1}\else
  \providecommand{\doi}{doi: \begingroup \urlstyle{rm}\Url}\fi

\bibitem[Barreno et~al.(2006)Barreno, Nelson, Sears, Joseph, and Tygar]{barreno2006can}
Barreno, M., Nelson, B., Sears, R., Joseph, A.~D., and Tygar, J.~D.
\newblock Can machine learning be secure?
\newblock In \emph{Proceedings of the 2006 ACM Symposium on Information, computer and communications security}, pp.\  16--25, 2006.

\bibitem[Barreno et~al.(2010)Barreno, Nelson, Joseph, and Tygar]{barreno2010security}
Barreno, M., Nelson, B., Joseph, A.~D., and Tygar, J.~D.
\newblock The security of machine learning.
\newblock \emph{Machine Learning}, 81:\penalty0 121--148, 2010.

\bibitem[Biggio et~al.(2012)Biggio, Nelson, and Laskov]{biggio2012poisoning}
Biggio, B., Nelson, B., and Laskov, P.
\newblock Poisoning attacks against support vector machines.
\newblock In \emph{Proc.~Int'l Conf.~Machine Learning}, pp.\  1467--1474, 2012.

\bibitem[Bozkurt et~al.(2021)Bozkurt, Esmaeili, Tristan, Brooks, Dy, and van~de Meent]{vae_rate}
Bozkurt, A., Esmaeili, B., Tristan, J.-B., Brooks, D., Dy, J., and van~de Meent, J.-W.
\newblock Rate-regularization and generalization in variational autoencoders.
\newblock In \emph{International Conference on Artificial Intelligence and Statistics}, pp.\  3880--3888. PMLR, 2021.

\bibitem[Chen et~al.(2023)Chen, Yuan, Cheng, Gong, Qin, Wang, and Huang]{sep}
Chen, S., Yuan, G., Cheng, X., Gong, Y., Qin, M., Wang, Y., and Huang, X.
\newblock Self-ensemble protection: Training checkpoints are good data protectors.
\newblock In \emph{Proc.~Int'l Conf.~Learning Representations}, 2023.

\bibitem[Deng et~al.(2009)Deng, Dong, Socher, Li, Li, and Fei-Fei]{imagenet}
Deng, J., Dong, W., Socher, R., Li, L.-J., Li, K., and Fei-Fei, L.
\newblock Imagenet: A large-scale hierarchical image database.
\newblock In \emph{Proc.~IEEE Int'l Conf.~Computer Vision and Pattern Recognition}, pp.\  248--255. Ieee, 2009.

\bibitem[Dolatabadi et~al.(2023)Dolatabadi, Erfani, and Leckie]{dolatabadi2023devil}
Dolatabadi, H.~M., Erfani, S., and Leckie, C.
\newblock The devil's advocate: Shattering the illusion of unexploitable data using diffusion models.
\newblock \emph{arXiv preprint arXiv:2303.08500}, 2023.

\bibitem[Feng et~al.(2019)Feng, Cai, and Zhou]{dc}
Feng, J., Cai, Q.-Z., and Zhou, Z.-H.
\newblock Learning to confuse: generating training time adversarial data with auto-encoder.
\newblock \emph{Proc.~Annual Conf.~Neural Information Processing Systems}, 32, 2019.

\bibitem[Fowl et~al.(2021)Fowl, Goldblum, Chiang, Geiping, Czaja, and Goldstein]{tap}
Fowl, L., Goldblum, M., Chiang, P.-y., Geiping, J., Czaja, W., and Goldstein, T.
\newblock Adversarial examples make strong poisons.
\newblock \emph{Proc.~Annual Conf.~Neural Information Processing Systems}, 34:\penalty0 30339--30351, 2021.

\bibitem[Fu et~al.(2022)Fu, He, Liu, Shen, and Tao]{rem}
Fu, S., He, F., Liu, Y., Shen, L., and Tao, D.
\newblock Robust unlearnable examples: Protecting data privacy against adversarial learning.
\newblock In \emph{Proc.~Int'l Conf.~Learning Representations}, 2022.

\bibitem[Goldblum et~al.(2022)Goldblum, Tsipras, Xie, Chen, Schwarzschild, Song, Madry, Li, and Goldstein]{goldblum2022dataset}
Goldblum, M., Tsipras, D., Xie, C., Chen, X., Schwarzschild, A., Song, D., Madry, A., Li, B., and Goldstein, T.
\newblock Dataset security for machine learning: Data poisoning, backdoor attacks, and defenses.
\newblock \emph{{IEEE} Trans. on Pattern Analysis and Machine Intelligence}, 45\penalty0 (2):\penalty0 1563--1580, 2022.

\bibitem[Goodfellow et~al.(2014)Goodfellow, Shlens, and Szegedy]{fgsm}
Goodfellow, I.~J., Shlens, J., and Szegedy, C.
\newblock Explaining and harnessing adversarial examples.
\newblock \emph{arXiv preprint arXiv:1412.6572}, 2014.

\bibitem[Gu et~al.(2017)Gu, Dolan-Gavitt, and Garg]{gu2017badnets}
Gu, T., Dolan-Gavitt, B., and Garg, S.
\newblock Badnets: Identifying vulnerabilities in the machine learning model supply chain.
\newblock \emph{arXiv preprint arXiv:1708.06733}, 2017.

\bibitem[Guo et~al.(2018)Guo, Rana, Cisse, and van~der Maaten]{guo2018countering}
Guo, C., Rana, M., Cisse, M., and van~der Maaten, L.
\newblock Countering adversarial images using input transformations.
\newblock In \emph{Proc.~Int'l Conf.~Learning Representations}, 2018.

\bibitem[He et~al.(2016)He, Zhang, Ren, and Sun]{resnet}
He, K., Zhang, X., Ren, S., and Sun, J.
\newblock Deep residual learning for image recognition.
\newblock In \emph{Proc.~IEEE Int'l Conf.~Computer Vision and Pattern Recognition}, pp.\  770--778, 2016.

\bibitem[Ho et~al.(2020)Ho, Jain, and Abbeel]{ho2020denoising}
Ho, J., Jain, A., and Abbeel, P.
\newblock Denoising diffusion probabilistic models.
\newblock \emph{Proc.~Annual Conf.~Neural Information Processing Systems}, pp.\  6840--6851, 2020.

\bibitem[Huang et~al.(2017)Huang, Liu, Van Der~Maaten, and Weinberger]{densenet}
Huang, G., Liu, Z., Van Der~Maaten, L., and Weinberger, K.~Q.
\newblock Densely connected convolutional networks.
\newblock In \emph{Proc.~IEEE Int'l Conf.~Computer Vision and Pattern Recognition}, pp.\  4700--4708, 2017.

\bibitem[Huang et~al.(2021)Huang, Ma, Erfani, Bailey, and Wang]{em}
Huang, H., Ma, X., Erfani, S.~M., Bailey, J., and Wang, Y.
\newblock Unlearnable examples: Making personal data unexploitable.
\newblock In \emph{Proc.~Int'l Conf.~Learning Representations}, 2021.

\bibitem[Jiang et~al.(2023)Jiang, Diao, Wang, Sun, Wang, and Hong]{jiang2023unlearnable}
Jiang, W., Diao, Y., Wang, H., Sun, J., Wang, M., and Hong, R.
\newblock Unlearnable examples give a false sense of security: Piercing through unexploitable data with learnable examples.
\newblock \emph{arXiv preprint arXiv:2305.09241}, 2023.

\bibitem[Kingma \& Welling(2014)Kingma and Welling]{vae}
Kingma, D.~P. and Welling, M.
\newblock Auto-encoding variational bayes.
\newblock In \emph{Proc.~Int'l Conf.~Learning Representations}, 2014.

\bibitem[Koh \& Liang(2017)Koh and Liang]{koh2017understanding}
Koh, P.~W. and Liang, P.
\newblock Understanding black-box predictions via influence functions.
\newblock In \emph{Proc.~Int'l Conf.~Machine Learning}, pp.\  1885--1894. PMLR, 2017.

\bibitem[Krizhevsky et~al.(2009)Krizhevsky, Hinton, et~al.]{cifar}
Krizhevsky, A., Hinton, G., et~al.
\newblock Learning multiple layers of features from tiny images.
\newblock 2009.

\bibitem[Lin et~al.(2024)Lin, Yu, Xia, Jiang, Wang, Yu, Liu, Fu, Wang, Tang, et~al.]{lin2024safeguarding}
Lin, X., Yu, Y., Xia, S., Jiang, J., Wang, H., Yu, Z., Liu, Y., Fu, Y., Wang, S., Tang, W., et~al.
\newblock Safeguarding medical image segmentation datasets against unauthorized training via contour-and texture-aware perturbations.
\newblock \emph{arXiv preprint arXiv:2403.14250}, 2024.

\bibitem[Liu et~al.(2023)Liu, Zhao, and Larson]{iss}
Liu, Z., Zhao, Z., and Larson, M.
\newblock Image shortcut squeezing: Countering perturbative availability poisons with compression.
\newblock \emph{Proc.~Int'l Conf.~Machine Learning}, 2023.

\bibitem[Lu et~al.(2023)Lu, Kamath, and Yu]{10.5555/3618408.3619357}
Lu, Y., Kamath, G., and Yu, Y.
\newblock Exploring the limits of model-targeted indiscriminate data poisoning attacks.
\newblock In \emph{Proc.~Int'l Conf.~Machine Learning}, 2023.

\bibitem[Madry et~al.(2018)Madry, Makelov, Schmidt, Tsipras, and Vladu]{pgd}
Madry, A., Makelov, A., Schmidt, L., Tsipras, D., and Vladu, A.
\newblock Towards deep learning models resistant to adversarial attacks.
\newblock In \emph{Proc.~Int'l Conf.~Learning Representations}, 2018.

\bibitem[Qin et~al.(2023{\natexlab{a}})Qin, Gao, Zhao, Ye, and Xu]{qin2023apbench}
Qin, T., Gao, X., Zhao, J., Ye, K., and Xu, C.-Z.
\newblock Apbench: A unified benchmark for availability poisoning attacks and defenses.
\newblock \emph{arXiv preprint arXiv:2308.03258}, 2023{\natexlab{a}}.

\bibitem[Qin et~al.(2023{\natexlab{b}})Qin, Gao, Zhao, Ye, and Xu]{qin2023learning}
Qin, T., Gao, X., Zhao, J., Ye, K., and Xu, C.-Z.
\newblock Learning the unlearnable: Adversarial augmentations suppress unlearnable example attacks.
\newblock \emph{arXiv preprint arXiv:2303.15127}, 2023{\natexlab{b}}.

\bibitem[Sandler et~al.(2018)Sandler, Howard, Zhu, Zhmoginov, and Chen]{mobilenetv2}
Sandler, M., Howard, A., Zhu, M., Zhmoginov, A., and Chen, L.-C.
\newblock Mobilenetv2: Inverted residuals and linear bottlenecks.
\newblock In \emph{Proc.~IEEE Int'l Conf.~Computer Vision and Pattern Recognition}, pp.\  4510--4520, 2018.

\bibitem[Sandoval-Segura et~al.(2022)Sandoval-Segura, Singla, Geiping, Goldblum, Goldstein, and Jacobs]{ar}
Sandoval-Segura, P., Singla, V., Geiping, J., Goldblum, M., Goldstein, T., and Jacobs, D.
\newblock Autoregressive perturbations for data poisoning.
\newblock \emph{Proc.~Annual Conf.~Neural Information Processing Systems}, 35:\penalty0 27374--27386, 2022.

\bibitem[Sandoval-Segura et~al.(2023)Sandoval-Segura, Singla, Geiping, Goldblum, and Goldstein]{lfu}
Sandoval-Segura, P., Singla, V., Geiping, J., Goldblum, M., and Goldstein, T.
\newblock What can we learn from unlearnable datasets?
\newblock In \emph{Proc.~Annual Conf.~Neural Information Processing Systems}, 2023.

\bibitem[Schmidhuber(2015)]{schmidhuber2015deep}
Schmidhuber, J.
\newblock Deep learning in neural networks: An overview.
\newblock \emph{Neural networks}, 61:\penalty0 85--117, 2015.

\bibitem[Schwarzschild et~al.(2021)Schwarzschild, Goldblum, Gupta, Dickerson, and Goldstein]{schwarzschild2021just}
Schwarzschild, A., Goldblum, M., Gupta, A., Dickerson, J.~P., and Goldstein, T.
\newblock Just how toxic is data poisoning? a unified benchmark for backdoor and data poisoning attacks.
\newblock In \emph{Proc.~Int'l Conf.~Machine Learning}, pp.\  9389--9398. PMLR, 2021.

\bibitem[Song et~al.(2021)Song, Sohl-Dickstein, Kingma, Kumar, Ermon, and Poole]{song2020score}
Song, Y., Sohl-Dickstein, J., Kingma, D.~P., Kumar, A., Ermon, S., and Poole, B.
\newblock Score-based generative modeling through stochastic differential equations.
\newblock In \emph{Proc.~Int'l Conf.~Learning Representations}, 2021.

\bibitem[Tao et~al.(2021)Tao, Feng, Yi, Huang, and Chen]{tao2021better}
Tao, L., Feng, L., Yi, J., Huang, S.-J., and Chen, S.
\newblock Better safe than sorry: Preventing delusive adversaries with adversarial training.
\newblock \emph{Proc.~Annual Conf.~Neural Information Processing Systems}, 34:\penalty0 16209--16225, 2021.

\bibitem[Wang et~al.(2024)Wang, Yu, Guo, and Wen]{wang2024benchmarking}
Wang, C., Yu, Y., Guo, L., and Wen, B.
\newblock Benchmarking adversarial robustness of image shadow removal with shadow-adaptive attacks.
\newblock In \emph{Proc.~IEEE Int'l Conf.~Acoustics, Speech, and Signal Processing}, pp.\  13126--13130. IEEE, 2024.

\bibitem[Wang et~al.(2018)Wang, Sun, Zhang, and Wang]{bdr}
Wang, J., Sun, J., Zhang, P., and Wang, X.
\newblock Detecting adversarial samples for deep neural networks through mutation testing.
\newblock \emph{arXiv preprint arXiv:1805.05010}, 2018.

\bibitem[Wen et~al.(2023)Wen, Zhao, Liu, Backes, Wang, and Zhang]{wen2022adversarial}
Wen, R., Zhao, Z., Liu, Z., Backes, M., Wang, T., and Zhang, Y.
\newblock Is adversarial training really a silver bullet for mitigating data poisoning?
\newblock In \emph{Proc.~Int'l Conf.~Learning Representations}, 2023.

\bibitem[Wu et~al.(2023)Wu, Chen, Xie, and Huang]{ops}
Wu, S., Chen, S., Xie, C., and Huang, X.
\newblock One-pixel shortcut: On the learning preference of deep neural networks.
\newblock In \emph{Proc.~Int'l Conf.~Learning Representations}, 2023.

\bibitem[Xia et~al.(2024)Xia, Yi, Jiang, and Ding]{xia2024mitigating}
Xia, S., Yi, Y., Jiang, X., and Ding, H.
\newblock Mitigating the curse of dimensionality for certified robustness via dual randomized smoothing.
\newblock In \emph{Proc.~Int'l Conf.~Learning Representations}, 2024.

\bibitem[Xiao et~al.(2015)Xiao, Biggio, Brown, Fumera, Eckert, and Roli]{xiao2015feature}
Xiao, H., Biggio, B., Brown, G., Fumera, G., Eckert, C., and Roli, F.
\newblock Is feature selection secure against training data poisoning?
\newblock In \emph{Proc.~Int'l Conf.~Machine Learning}, pp.\  1689--1698. PMLR, 2015.

\bibitem[Yu et~al.(2022{\natexlab{a}})Yu, Zhang, Chen, Yin, and Liu]{lsp}
Yu, D., Zhang, H., Chen, W., Yin, J., and Liu, T.-Y.
\newblock Availability attacks create shortcuts.
\newblock In \emph{Proceedings of the 28th ACM SIGKDD Conference on Knowledge Discovery and Data Mining}, pp.\  2367--2376, 2022{\natexlab{a}}.

\bibitem[Yu et~al.(2022{\natexlab{b}})Yu, Yang, Tan, and Kot]{yu2022towards}
Yu, Y., Yang, W., Tan, Y.-P., and Kot, A.~C.
\newblock Towards robust rain removal against adversarial attacks: A comprehensive benchmark analysis and beyond.
\newblock In \emph{Proc.~IEEE Int'l Conf.~Computer Vision and Pattern Recognition}, pp.\  6013--6022, 2022{\natexlab{b}}.

\bibitem[Yu et~al.(2023)Yu, Wang, Yang, Lu, Tan, and Kot]{yu2023backdoor}
Yu, Y., Wang, Y., Yang, W., Lu, S., Tan, Y.-P., and Kot, A.~C.
\newblock Backdoor attacks against deep image compression via adaptive frequency trigger.
\newblock In \emph{Proc.~IEEE Int'l Conf.~Computer Vision and Pattern Recognition}, pp.\  12250--12259, 2023.

\bibitem[Yuan \& Wu(2021)Yuan and Wu]{ntga}
Yuan, C.-H. and Wu, S.-H.
\newblock Neural tangent generalization attacks.
\newblock In \emph{Proc.~Int'l Conf.~Machine Learning}, pp.\  12230--12240. PMLR, 2021.

\bibitem[Zha et~al.(2023)Zha, Bhat, Lai, Yang, and Hu]{zha2023data}
Zha, D., Bhat, Z.~P., Lai, K.-H., Yang, F., and Hu, X.
\newblock Data-centric ai: Perspectives and challenges.
\newblock In \emph{Proceedings of the 2023 SIAM International Conference on Data Mining (SDM)}, pp.\  945--948. SIAM, 2023.

\bibitem[Zhang et~al.(2018)Zhang, Cisse, Dauphin, and Lopez-Paz]{zhang2018mixup}
Zhang, H., Cisse, M., Dauphin, Y.~N., and Lopez-Paz, D.
\newblock mixup: Beyond empirical risk minimization.
\newblock In \emph{Proc.~Int'l Conf.~Learning Representations}, 2018.

\bibitem[Zhao \& Lao(2022)Zhao and Lao]{zhao2022clpa}
Zhao, B. and Lao, Y.
\newblock Clpa: Clean-label poisoning availability attacks using generative adversarial nets.
\newblock In \emph{Proceedings of the AAAI Conference on Artificial Intelligence}, volume~36, pp.\  9162--9170, 2022.

\bibitem[Zhao et~al.(2017)Zhao, An, Gao, and Zhang]{zhao2017efficient}
Zhao, M., An, B., Gao, W., and Zhang, T.
\newblock Efficient label contamination attacks against black-box learning models.
\newblock In \emph{Proceedings of the Twenty-Sixth International Joint Conference on Artificial Intelligence}, pp.\  3945--3951, 2017.

\end{thebibliography}
\bibliographystyle{icml2024}

\newpage
\appendix
\onecolumn
\section{Proofs}
In this section, we provide the proofs of our theoretical results in Section~\ref{vae_explaination}.
\subsection{Proof of Proposition~\ref{proposition1}} \label{Proof_of_Proposition1}
Consider the following binary classification problem with regards to the features extracted from the data ${\boldsymbol{v}} = ({\boldsymbol{v}}_c, {\boldsymbol{v}}_s^t)$ consisting of a predictive feature $x_c$ of a Gaussian mixture ${\mathcal{G}}_c$ and a non-predictive feature ${\boldsymbol{v}}_s^t$ which follows:
\begin{equation}
    y \stackrel{{u \cdot a\cdot r}}{\sim} \{0, 1\}, ~{\boldsymbol{v}}_c~{\sim}~\mathcal{N}({\boldsymbol{\mu}}_{c}^{y},{\boldsymbol{\Sigma}}_c), ~{\boldsymbol{v}}_s^{t}~{\sim}~\mathcal{N}({\boldsymbol{\mu}}^{t},{\boldsymbol{\Sigma}}^t), ~{\boldsymbol{v}}_c \indep {\boldsymbol{v}}_s^t, \quad \Pr(y=0)=\Pr(y=1).
    \label{distribution_appendix}
\end{equation}

\noindent \textbf{Proposition~\ref{proposition1} (restated)} \textit{
For the data ${\boldsymbol{v}} = ({\boldsymbol{v}}_c, {\boldsymbol{v}}_s^t)$ following the distribution~(\ref{distribution_appendix}), the optimal separating hyperplane using a Bayes classifier is formulated by:}
\begin{equation}
    \boldsymbol{w}_c^{\top}({\boldsymbol{v}}_c^*-\frac{\boldsymbol{\mu}_c^0+\boldsymbol{\mu}_c^1}{2})=0, \quad \text{where} ~\boldsymbol{w}_c = \boldsymbol{\Sigma}_c^{-1}(\boldsymbol{\mu}_c^0-\boldsymbol{\mu}_c^1).
    \label{hyperplane_appendix}
\end{equation}

\textit{Proof.} Given ${\boldsymbol{v}} = ({\boldsymbol{v}}_c, {\boldsymbol{v}}_s^t)$ following the distribution~(\ref{distribution_appendix}), the optimal decision rule is the maximum a-posteriori probability rule for a Bayes classifier:
\begin{equation}
    \begin{split}
        i^{*}({\boldsymbol{v}}) &= \argmax_i\Pr(y=i|{\boldsymbol{v}})\\
        & = \argmax_i\big[\Pr(y=i)\Pr({\boldsymbol{v}}|y=i)\big]\\
        & = \argmax_i\big[\ln \Pr({\boldsymbol{v}}|y=i)\big]\\
        & = \argmax_i\big[\ln \Pr({\boldsymbol{v}}_c|y=i)+\ln \Pr({\boldsymbol{v}}_s^t|y=i)\big]\\
        & = \argmax_i\big[\ln \Pr({\boldsymbol{v}}_c|y=i)\big]\\
        &=\argmax_i\Big[\ln\big[(2\pi)^{-\frac{D}{2}}|\boldsymbol{\Sigma}_c|^{-\frac{1}{2}}\exp(-\frac{1}{2}({\boldsymbol{v}}_c-\boldsymbol{\mu}_c^i)^\top\boldsymbol{\Sigma}_c^{-1}({\boldsymbol{v}}_c-\boldsymbol{\mu}_c^i))\big]\Big]\\
        &=\argmin_i\big[({\boldsymbol{v}}_c-\boldsymbol{\mu}_c^i)^\top\boldsymbol{\Sigma}_c^{-1}({\boldsymbol{v}}_c-\boldsymbol{\mu}_c^i))\big]\\
        &=\argmin_i\big[{\boldsymbol{v}}_c^\top\boldsymbol{\Sigma}_c^{-1}{\boldsymbol{v}}_c - 2 \mu_c^i{^{\top}}\boldsymbol{\Sigma}_c^{-1}{\boldsymbol{v}}_c + \boldsymbol{\mu}_c^i{^{\top}}\boldsymbol{\Sigma}_c^{-1}\mu_c^i\big]\\
        &=\argmax_i\big[\boldsymbol{\mu}_c^i{^{\top}}\boldsymbol{\Sigma}_c^{-1}{\boldsymbol{v}}_c - \frac{1}{2} \boldsymbol{\mu}_c^i{^{\top}}\boldsymbol{\Sigma}_c^{-1}\boldsymbol{\mu}_c^i\big]
    \end{split},
\end{equation}
where $D$ is the dimensions. Thus, the hyperplane is formulated by:
\begin{equation}
    \begin{split}
    &\boldsymbol{\mu}_c^0{^{\top}}\boldsymbol{\Sigma}_c^{-1}{\boldsymbol{v}}_c - \frac{1}{2} \boldsymbol{\mu}_c^0{^{\top}}\boldsymbol{\Sigma}_c^{-1}\boldsymbol{\mu}_c^0 = \boldsymbol{\mu}_c^1{^{\top}}\boldsymbol{\Sigma}_c^{-1}{\boldsymbol{v}}_c - \frac{1}{2} \boldsymbol{\mu}_c^1{^{\top}}\boldsymbol{\Sigma}_c^{-1}\boldsymbol{\mu}_c^1\\
    &\iff  \boldsymbol{w}_c^{\top}({\boldsymbol{v}}_c^*-\frac{\boldsymbol{\mu}_c^0+\boldsymbol{\mu}_c^1}{2})=0, \quad \text{where} ~\boldsymbol{w}_c = \boldsymbol{\Sigma}_c^{-1}(\boldsymbol{\mu}_c^0-\boldsymbol{\mu}_c^1).
    \end{split}
\end{equation}
\subsection{Proof of Theorem~\ref{theorem1}}\label{Proof_of_Theorem1}
We assume that a malicious attacker modifies ${\boldsymbol{v}}_s^t$ to ${\boldsymbol{v}}_s$ of the following distributions ${\mathcal{G}}_s$ to make it predictive for training a Bayes classifier:
\begin{equation}
    y \stackrel{{u \cdot a\cdot r}}{\sim} \{0, 1\}, \quad {\boldsymbol{v}}_s~{\sim}~\mathcal{N}({\boldsymbol{\mu}}_{s}^{y},{\boldsymbol{\Sigma}}_s), \quad {\boldsymbol{v}}_c \indep {\boldsymbol{v}}_s.
    \label{poison_distribution_appendix}
\end{equation}

\noindent \textbf{Theorem~\ref{theorem1} (restated)} \textit{
Consider the training data for the Bayes classifier is modified from ${\boldsymbol{v}} = ({\boldsymbol{v}}_c, {\boldsymbol{v}}_s^t)$ in Eq.~\ref{distribution_appendix} to ${\boldsymbol{v}} = ({\boldsymbol{v}}_c, {\boldsymbol{v}}_s)$ in Eq.~\ref{poison_distribution_appendix}, the hyperplane is shifted with a distance given by}
\begin{equation}
    d = \frac{{\lVert \boldsymbol{w}_s^{\top}({\boldsymbol{v}}_s-\frac{\boldsymbol{\mu}_s^0+\boldsymbol{\mu}_s^1}{2}) \rVert}_2}{{\lVert \boldsymbol{w}_c \rVert}_2}
    , \quad \text{where} ~\boldsymbol{w}_c = \boldsymbol{\Sigma}_c^{-1}(\boldsymbol{\mu}_c^0-\boldsymbol{\mu}_c^1), ~\boldsymbol{w}_s = \boldsymbol{\Sigma}_s^{-1}(\boldsymbol{\mu}_s^0-\boldsymbol{\mu}_s^1).
\end{equation}

\textit{Proof.} After modifying ${\boldsymbol{v}}_s^t$ to ${\boldsymbol{v}}_s$, the learned separating hyperplane on the poisoned distributions ${\mathcal{G}}_p$ = (${\mathcal{G}}_c$, ${\mathcal{G}}_s$) turns to (following Proposition~\ref{proposition1}):
\begin{equation}
\begin{split}
    w^{\top}&(\left[\begin{matrix}
{\boldsymbol{v}}_c^* -\frac{\boldsymbol{\mu}_c^0+\boldsymbol{\mu}_c^1}{2}\\
{\boldsymbol{v}}_s -\frac{\boldsymbol{\mu}_s^0+\boldsymbol{\mu}_s^1}{2}\\
\end{matrix} \right])=0 \iff 
\boldsymbol{w}_c^{\top}({\boldsymbol{v}}_c^*-\frac{\boldsymbol{\mu}_c^0+\boldsymbol{\mu}_c^1}{2}) = -\boldsymbol{w}_s^{\top}({\boldsymbol{v}}_s-\frac{\boldsymbol{\mu}_s^0+\boldsymbol{\mu}_s^1}{2}),
\\
&\text{where} ~\boldsymbol{w} = \left[\begin{matrix}
\boldsymbol{\Sigma}_c^{-1} & 0\\
0 & \boldsymbol{\Sigma}_s^{-1}\\
\end{matrix} \right]\left[\begin{matrix}
\boldsymbol{\mu}_c^0-\boldsymbol{\mu}_c^1\\
\boldsymbol{\mu}_s^0-\boldsymbol{\mu}_s^1\\
\end{matrix} \right]=\left[\begin{matrix}
\boldsymbol{w}_c\\
\boldsymbol{w}_s\\
\end{matrix} \right], ~\boldsymbol{w}_c = \boldsymbol{\Sigma}_c^{-1}(\boldsymbol{\mu}_c^0-\boldsymbol{\mu}_c^1).
\end{split}
\end{equation}

Thus, compared to the original hyperplane as stated in Eq.~\ref{hyperplane_appendix}, the hyperplane on the poisoned distribution is shifted with a distance $d$:
\begin{equation}
    d = \frac{{\lVert \boldsymbol{w}_s^{\top}({\boldsymbol{v}}_s-\frac{\boldsymbol{\mu}_s^0+\boldsymbol{\mu}_s^1}{2}) \rVert}_2}{{\lVert \boldsymbol{w}_c \rVert}_2}
    \label{hyperplane_shifted_equation_appendix}
\end{equation}
When conducting evaluations on the testing data that follows the same distribution as the clean data ${\boldsymbol{v}} = ({\boldsymbol{v}}_c, {\boldsymbol{v}}_s^t)$, with the term ${\boldsymbol{v}}_s$ in Eq.~\ref{hyperplane_shifted_equation_appendix} replaced by ${\boldsymbol{v}}_s^t$, the shifted distance $d$ is given by
\begin{equation}
    d = \frac{{\lVert \boldsymbol{w}_s^{\top}({\boldsymbol{v}}_s^t-\frac{\boldsymbol{\mu}_s^0+\boldsymbol{\mu}_s^1}{2}) \rVert}_2}{{\lVert \boldsymbol{w}_c \rVert}_2} \propto \frac{{\lVert \boldsymbol{w}_s \rVert}_2}{{\lVert \boldsymbol{w}_c \rVert}_2}.
\end{equation}
And it leads to a greater prediction error if ${{\lVert \boldsymbol{w}_s \rVert}_2} \gg {{\lVert \boldsymbol{w}_c \rVert}_2}$.

\subsection{Proof of Theorem~\ref{theorem2}}\label{Proof_of_Theorem2}
Consider a variable ${\boldsymbol{v}} = ({v}_1, \dots, {v}_d)$ following a mixture of two Gaussian distributions ${\mathcal{G}}$: 
\begin{equation}
\begin{split}
    y \stackrel{{u \cdot a\cdot r}}{\sim} &\{0, 1\}, \quad {\boldsymbol{v}}~{\sim}~\mathcal{N}({\boldsymbol{\mu}}^{y},{\boldsymbol{\Sigma}}), \quad x_i \indep x_j, 
 \quad \Pr(y=0)=\Pr(y=1), \\
  &{v}_i~{\sim}~\mathcal{N}({\mu}_i^{y},{\sigma_i}), \quad p_{{v}_i}({v}) = \frac{\mathcal{N}({v};{\mu}_i^{0},{\sigma_i}) + \mathcal{N}({v};{\mu}_i^{1},{\sigma_i})}{2}.
  \end{split}
\end{equation}
Each dimensional feature ${v}_i$ is also modeled as a Gaussian mixture. 
To start, we normalize each feature through a linear operation to achieve a distribution with zero mean and unit variance. Firstly, we calculate the mean and standard deviation of ${v}_i$:
\begin{equation}
    \hat{\mu}_i = \mathbb{E}_{{v}_i}\big[{v}_i\big]=\frac{{\mu}^{0}_i+{\mu}^{1}_i}{2}, \quad \text{Var}[{v}_i] = \mathbb{E}_{{v}_i}\big[({v}_i)^2\big] - \mathbb{E}_{{v}_i}\big[{v}_i\big]^2=\sigma_i^2 + (\frac{{\mu}^{0}_i-{\mu}^{1}_i}{2})^2.
\end{equation}
Thus, the linear operation and the modified density function can be expressed as follows:
\begin{equation}
\begin{split}
    &z_i = \frac{{v}_i - \hat{\mu}_i}{\sqrt{{(\sigma_i)}^2+{(\delta_i)}^2}}, ~p_{z_i}({v}) = \frac{p_0({v})+p_1({v})}{2}, ~p_0({v}) = \mathcal{N}({v};-\hat{\delta}_i,\hat{\sigma}_i), ~ p_1({v}) = \mathcal{N}({v};\hat{\delta}_i,\hat{\sigma}_i)\\
    &~\text{where} ~\hat{\mu}_i =\frac{{\mu}^{0}_i+{\mu}^{1}_i}{2}, ~\delta_i =|\frac{{\mu}^{0}_i-{\mu}^{1}_i}{2}|, ~\hat{\delta}_i=\delta_i/\sqrt{{(\sigma_i)}^2+{(\delta_i)}^2}, ~\hat{\sigma}_i=\sigma_i/\sqrt{{(\sigma_i)}^2+{(\delta_i)}^2}.
\end{split}
\end{equation}

\noindent \textbf{Theorem~\ref{theorem2} (restated)} \textit{
    Denote $r=\frac{\delta_i}{\sigma_i}>0$, the Kullback–Leibler divergence between $p_{z_i}({v})$ and a standard normal distribution $\mathcal{N}({v};0,1)$ is tightly bounded by}
\begin{equation}
    \frac{1}{2}\ln{(1+r^2)}-\ln2 \leq D_{KL}(p_{z_i}({v})\Vert\mathcal{N}({v};0,1)) \leq \frac{1}{2}\ln{(1+r^2)}.
\end{equation}
 \textit{
and observes the following property }
\begin{equation}
\begin{split}
    \uparrow r_i & \quad \implies \quad \uparrow S(r_i)=D_{KL}(p_{z_i}({v})\Vert\mathcal{N}({v};0,1)).
\end{split}
\end{equation}

\textit{Proof.} We estimate the Kullback–Leibler divergence between $p_{z_i}({v})$ and $\mathcal{N}({v};0,1)$:
\begin{equation}
\begin{split}
    D_{KL}(p_{z_i}({v})\Vert\mathcal{N}({v};0,1)) &= \int_{-\infty}^{\infty}p_{z_i}({v})\ln\frac{p_{z_i}({v})}{\mathcal{N}({v};0,1)}d{v}\\
    &=-H(p_{z_i}({v}))+H((p_{z_i}({v}),\mathcal{N}({v};0,1)))\\
    &=-H(p_{z_i}({v}))+\frac{1}{2}(1+\ln2\pi),\\
\end{split}
\end{equation}

As the funcion $\mathcal{N}({v};0,1)))$ is given by
\begin{equation}
    \mathcal{N}({v};0,1))) = \frac{1}{\sqrt{2\pi}}e^{-\frac{{v}^2}{2}},
\end{equation}
then the term $H((p_{z_i}({v}),\mathcal{N}({v};0,1)))$ can be formulated as
\begin{equation}
    \begin{split}
        H((p_{z_i}({v}),\mathcal{N}({v};0,1)))&=\int_{-\infty}^{\infty}p_{z_i}({v})\ln\frac{1}{\mathcal{N}({v};0,1)}d{v}\\
        &=\int_{-\infty}^{\infty}p_{z_i}({v})\big[\frac{1}{2}\ln2\pi + \frac{1}{2}{v}^2\big]d{v}\\
        &=\frac{1}{2}\ln2\pi\int_{-\infty}^{\infty}p_{z_i}({v})d{v} + \frac{1}{2}\int_{-\infty}^{\infty}{v}^2p_{z_i}({v})d{v}\\
        &=\frac{1}{2}\ln2\pi + \frac{1}{2}\int_{-\infty}^{\infty}{v}^2\frac{p_0({v})+p_1({v})}{2}d{v}\\
        & = \frac{1}{2}\ln2\pi + \frac{1}{4}\big[{E}_{{v} \sim p_0({v})}[{v}^2] + {E}_{{v} \sim p_1({v})}[{v}^2]\big]\\
        & = \frac{1}{2}\ln2\pi + \frac{1}{4}\big[{E}_{p_0({v})}[{v}]^2 + {Var}_{p_0({v})}[{v}] + {E}_{p_1({v})}[{v}]^2 + {Var}_{p_1({v})}[{v}]\big]\\
        & =  \frac{1}{2}\ln2\pi +\frac{1}{2}
    \end{split}
\end{equation}

As the entropy ${H}(p)$ is concave in the probability mass function $p$, a lower bound of  ${H}(p_{z_i})$ is given by:
\begin{equation}
\begin{split}
   {H}(p_{z_i}({v})) & = H(\frac{p_0({v})+p_1({v})}{2})\\
   &\geq \frac{1}{2}{H}(p_0({v})) + \frac{1}{2}{H}(p_1({v}))\\
   &= \frac{1}{2}{H}(\mathcal{N}({v};-\frac{r}{\sqrt{1+{r^2}}},\frac{1}{\sqrt{1+{r^2}}}) + \frac{1}{2}{H}(\mathcal{N}({v};\frac{r}{\sqrt{1+{r^2}}},\frac{1}{\sqrt{1+{r^2}}})\\
   & = \frac{1}{2}\Big[\frac{1}{2}(1+\ln(2\pi(\frac{1}{\sqrt{1+{r^2}}})^2)) + \frac{1}{2}(1+\ln(2\pi(\frac{1}{\sqrt{1+{r^2}}})^2))\Big]\\
    &=\frac{1}{2}(1+\ln2\pi) - \frac{1}{2}\ln{(1+r^2)}.\\
\end{split}
\end{equation}
The upper bound of  ${H}(p_{z_i})$ is given by:
\begin{equation}
\begin{split}
    &{H}(p_{z_i}) = -\int_{-\infty}^{\infty}p_{z_i}({v})\ln{p_{z_i}({v})}d{v}\\
    &= -\int_{-\infty}^{\infty}\frac{p_o({v})+p_1({v})}{2}\ln{\frac{p_o({v})+p_1({v})}{2}}d{v}\\
    &= -\frac{1}2\Big[\int_{-\infty}^{\infty}p_o({v})[\ln{\frac{p_o({v})}{2}+\ln(1+\frac{p_1({v})}{p_0({v})})}]d{v} +\int_{-\infty}^{\infty}p_1({v})[\ln\frac{p_1({v})}{2}+\ln(1+\frac{p_0({v})}{p_1({v})})]d{v}\Big]\\
    &\leq -\frac{1}2\Big[\int_{-\infty}^{\infty}p_o({v})\ln\frac{p_o({v})}{2}d{v} +\int_{-\infty}^{\infty}p_1({v})\ln{\frac{p_1({v})}{2}}d{v}\Big]\\
    &= \frac{1}{2}\Big[H(p_0)+H(p_1)+2\ln2\Big]\\
    &= \frac{1}{2}(1+\ln2\pi) - \frac{1}{2}\ln{(1+r^2)}+\ln2.
\end{split}
\end{equation}
Thus, the Kullback–Leibler divergence is bounded by :
\begin{equation}
    \frac{1}{2}\ln{(1+r^2)}-\ln2 \leq D_{KL}(p_{z_i}({v})\Vert\mathcal{N}({v};0,1)) \leq \frac{1}{2}\ln{(1+r^2)}.
\end{equation}
Since the lower and upper bounds differ by a constant term, and the lower bound increases significantly as $r$ rises, the Kullback–Leibler divergence is asymptotically tightly bounded by:
\begin{equation}
    D_{KL}(p_{z_i}({v})\Vert\mathcal{N}({v};0,1)) = \Theta (\ln(1+r^2))
\end{equation}

\subsection{Proof of Proposition~\ref{proposition2}}\label{Proof_of_Proposition2}
\noindent \textbf{Proposition~\ref{proposition2} (restated)} \textit{
    The conditional entropy of a Gaussian mixture ${\boldsymbol{v}}_s$ of ${\mathcal{G}}_s$ in Eq.~\ref{poison_distribution_appendix} is given by}
\begin{equation}
    H({\boldsymbol{v}}_s|y_i) = \frac{D}{2}(1+\ln(2\pi)) + \frac{1}{2}\ln|\boldsymbol{\Sigma}_s|,
\end{equation}
\textit{where} $D$ \textit{is the dimensions of the features.} \textit{If each feature ${v}_s^d$ is independent, then}
\begin{equation}
    H({\boldsymbol{v}}_s|y_i) =\frac{D}{2}(1+\ln(2\pi)) + \sum_{d=1}^{D} \ln\sigma_s^d.
\end{equation}
\textit{Proof.} For the variable follows a Gaussian distribution:
\begin{equation}
    {\boldsymbol{v}}~{\sim}~\mathcal{N}_D({\boldsymbol{\mu}},{\boldsymbol{\Sigma}}), 
\end{equation}
The derivation of its entropy is given by
\begin{equation}
\begin{split}
    H({\boldsymbol{v}}) &= - \int p({\boldsymbol{v}})\ln p({\boldsymbol{v}})d{\boldsymbol{v}}\\
    &=-\mathbb{E}\big[\ln\mathcal{N}_D({\boldsymbol{\mu}},{\boldsymbol{\Sigma}})\big]\\
    & = -\mathbb{E}\Big[\ln\big[(2\pi)^{-\frac{D}{2}}|\boldsymbol{\Sigma}|^{-\frac{1}{2}}\exp(-\frac{1}{2}({\boldsymbol{v}}-\boldsymbol{\mu})^\top\boldsymbol{\Sigma}^{-1}({\boldsymbol{v}}-\boldsymbol{\mu}))\big]\Big]\\
    & = \frac{D}{2}\ln(2\pi) + \frac{1}{2}\ln|\boldsymbol{\Sigma}|+\frac{1}{2}\mathbb{E}\big[({\boldsymbol{v}}-\boldsymbol{\mu})^\top\boldsymbol{\Sigma}^{-1}({\boldsymbol{v}}-\boldsymbol{\mu}))\big]\\
    & \overset{*}{=} \frac{D}{2}(1+\ln(2\pi)) + \frac{1}{2}\ln|\boldsymbol{\Sigma}|
\end{split}
\end{equation}
Step $*$ is a little trickier. It relies on several properties of the trace operator:
\begin{equation}
\begin{split}
\mathbb{E}\big[({\boldsymbol{v}}-\boldsymbol{\mu})^\top\boldsymbol{\Sigma}^{-1}({\boldsymbol{v}}-\boldsymbol{\mu}))\big] &= \mathbb{E}\Big[\text{tr}\big[({\boldsymbol{v}}-\boldsymbol{\mu})^\top\boldsymbol{\Sigma}^{-1}({\boldsymbol{v}}-\boldsymbol{\mu}))\big]\Big]\\
& = \mathbb{E}\Big[\text{tr}\big[\boldsymbol{\Sigma}^{-1}({\boldsymbol{v}}-\boldsymbol{\mu})({\boldsymbol{v}}-\boldsymbol{\mu})^\top\big]\Big]\\
& = \text{tr}\Big[\boldsymbol{\Sigma}^{-1}\mathbb{E}\big[({\boldsymbol{v}}-\boldsymbol{\mu})({\boldsymbol{v}}-\boldsymbol{\mu})^\top\big]\Big]\\
& = \text{tr}\Big[\boldsymbol{\Sigma}^{-1}\boldsymbol{\Sigma}\Big]\\
& = \text{tr}(\boldsymbol{I}_D)\\
& = \frac{D}{2}
\end{split}
\end{equation}

\newpage
\section{Detailed implementation} 
\subsection{KLD Loss}\label{kld_implementation}
For the implementation of KLD loss in Eq.~\ref{rate_constrained loss} and Eq.~\ref{main_loss}, we follows the widely-used version from \citet{vae}. The detailed loss formulation is given
\begin{equation}
\begin{split}
    &\text{KLD}(\boldsymbol{z},\mathcal{N}(0,\boldsymbol{I})) = -\frac{1}{2} \sum_{j=1}^{J}(1+\log(\sigma_j)^2-(\mu_j)^2-(\sigma_j)^2),\\
    &\text{where}~~\boldsymbol{z} = \boldsymbol{\mu}+\boldsymbol{\sigma}\odot\boldsymbol{\epsilon}, ~\text{and} ~~\boldsymbol{\epsilon} \sim \mathcal{N}(0,\boldsymbol{I}).
\end{split}    
\end{equation}

\subsection{D-VAE}
In the implementation of D-VAE, the encoder comprises 7 convolutional layers with Batch Normalization, while the decoder for both branches consists of 4 convolutional layers with Instance Normalization. To predict the mean $\mu$ and standard deviation $\sigma$, we employ one convolutional layer with a kernel size of 1 for each variable.

During the training of D-VAE, we configure the number of training epochs to be 60 for CIFAR-10 and CIFAR-100. However, for ImageNet, which involves significant computational demands, we limit the training epochs to 20. It's important to note that we do not use any transformations on the training data when training D-VAEs.
For D-VAE training on unlearnable CIFAR-10/100, we use a KLD target of 1.0 in the first stage and 3.0 in the second stage, with only a single $\times 0.5$ downsampling to preserve image quality. 
For ImageNet, which has higher-resolution images, we employ more substantial downsampling ($\times 0.125$) in the first stage and set a KLD target of 1.5, while the second stage remains the same as with CIFAR. 
When comparing the unlearnable input and the reconstructed output, these hyperparameters yield PSNRs of around 28 for CIFAR and 30 for ImageNet.

\section{Disentangled perturbations}\label{disentangled_poison_patterns}
Given that the defender lacks groundtruth values for the perturbations $\boldsymbol{p}$, it is not possible to optimize $\boldsymbol{u_y}$ and $D_{\theta_p}$ to learn to predict $\boldsymbol{\hat{p}}$ directly by minimizing ${\lVert \boldsymbol{p} - \boldsymbol{\hat{p}} \rVert}_2^2$ during model training. Instead, as the residuals $\boldsymbol{x} - \boldsymbol{\hat{x}}$ contain the majority of the groundtruth $\boldsymbol{p}$ when imposing a low target value on the KLD loss, we propose minimizing ${\lVert (\boldsymbol{x} - \boldsymbol{\hat{x}}) - \boldsymbol{\hat{p}} \rVert}_2^2$. As $\boldsymbol{\hat{p}}$ is generated by $\boldsymbol{u_y}+\boldsymbol{z}$, which has an information bottleneck, it is hard to achieve a perfect reconstruction of $\boldsymbol{p}$, and $\boldsymbol{\hat{p}}$ is most likely to be a part of $\boldsymbol{p}$. In Table~\ref{dis_norm}, we offer the $\ell_2$-norm of both $\boldsymbol{p}$ and $\boldsymbol{\hat{p}}$ , and we can see that the $\boldsymbol{\hat{p}}$ has a smaller amplitude. In Section 4.2, the experiments show that the $\boldsymbol{\hat{p}}$ remains effective as poisoning patterns. Notably, the amplitude of $\boldsymbol{\hat{p}}$ is comparable to that of $\boldsymbol{{p}}$, with $\boldsymbol{\hat{p}}$ being slightly smaller than $\boldsymbol{{p}}$ except for OPS. 

The visual results of the normalized perturbations can be seen in Figure~\ref{dvae}, and we observe the visually similarity between $\boldsymbol{\hat{p}}$ and $\boldsymbol{{p}}$, especially for LSP and OPS. Additionally, since LSP and OPS use class-wise perturbations (i.e., perturbations are identical for each class of images), they exhibit lower class-conditioned entropy compared to other attack methods that employ sample-wise perturbations. This makes the reconstruction of LSP and OPS perturbations much easier.
\begin{table}[t]
    \centering
\setlength\tabcolsep{7.0pt}
    \caption{Evaluation on the amplitude of the disentangled $\boldsymbol{\hat{p}}$ and groundtruth $\boldsymbol{{p}}$.  }
    \centering
    \scalebox{1.0}{
    \begin{tabular}{ c | c | c  c  c  c  c  c }
    \toprule
     \multirow{1}*{Datasets} & Test Set& EM & REM & NTGA & LSP & AR & OPS \\
    \midrule
     \multirow{3}*{CIFAR-10} &$\Vert \boldsymbol{p} \Vert_2$&1.53 & 1.80 & 2.96 & 0.99 & 0.98 & 1.27\\
     &$\Vert \boldsymbol{\hat{p}} \Vert_2$&1.24 & 0.92 & 0.71 & 0.73 & 0.68 & 1.77\\
     &MSE($\boldsymbol{p}$, $\boldsymbol{\hat{p}}$) / $10^{-4}$&4.4&6.6&12&3.5&5.2&2.3\\
     &PSNR($\boldsymbol{p}$, $\boldsymbol{\hat{p}}$) / dB&33.6&31.8&29.3&34.7&32.9&36.5\\
     \midrule
     \multirow{3}*{CIFAR-100}&$\Vert \boldsymbol{p} \Vert_2$&1.34 & 1.73 &  -   & 0.99 & 0.98 & 1.34 \\
     & $\Vert \boldsymbol{\hat{p}} \Vert_2$&0.81 & 0.69 &  -   & 0.69 & 0.78 & 1.82\\
     &MSE($\boldsymbol{p}$, $\boldsymbol{\hat{p}}$) / $10^{-4}$&4.1&7.1&-&4.1&4.7&2.0\\
    & PSNR($\boldsymbol{p}$, $\boldsymbol{\hat{p}}$) / dB&33.9&31.5&-&34.0&33.1&37.3\\
    \bottomrule
    \end{tabular}
    \label{dis_norm}}
\end{table}

\section{More results on partial poisoning}\label{more_partial_poisoning}
In Table~\ref{more_detection_poison_generation} and Table~\ref{more_partial}, we provide additional results on detecting UEs, increasing amounts of UEs, and experimental results on UEs purification in the partial poisoning settings.
\begin{table}[t]\small
\parbox{.6\linewidth}{
  \centering
  \caption{Performance of detecting UEs or increasing UEs with various poison ratios on CIFAR-10.  }
  \scalebox{1.0}{
  \setlength\tabcolsep{5.0pt}
    \centering
 \begin{tabular}{ c | c | c  c  c  c || c | c }
    \toprule
     Task& \multicolumn{5}{c||}{Detecting UEs} & \multicolumn{2}{c}{Increasing UEs}\\
    \midrule
     {UEs} & \multirow{1}*{Ratio} &  Acc. & Recall & Precision & F1-score & Ratio & Test Acc.\\
    \midrule
     EM&\multirow{4}*{0.2} &0.918&1.0&0.709 & 0.830&\multirow{4}*{0.01} & 0.1009 \\
     REM&&0.561
&1.0&0.312 & 0.476& & 0.2900\\
     LSP&&0.777
&1.0&0.472 & 0.641& & 0.1558\\
OPS&&0.724
&0.993&0.420 & 0.590& & 0.2059\\
     \midrule
     EM&\multirow{4}*{0.4} &0.939&1.0&0.869 & 0.930&\multirow{4}*{0.02}& 0.1011\\
     REM&&0.785
&1.0&0.651 & 0.789& & 0.2777 \\
     LSP&&0.905
&1.0&0.807 & 0.893& & 0.1633\\
OPS&&0.842
&0.991&0.719 & 0.833& & 0.2015\\
\midrule
     EM&\multirow{4}*{0.6} &0.961&1.0&0.938 & 0.968&\multirow{4}*{0.04} & 0.1229\\
     REM&&0.909
&0.999&0.868 & 0.930 && 0.2319\\
     LSP&&0.941
&0.999&0.912 &  0.954& & 0.1405 \\
OPS&&0.910
&0.993&0.874 & 0.930 && 0.1632\\
\midrule
EM&\multirow{4}*{0.8} &0.982&1.0&0.978 & 0.989&\multirow{4}*{0.08} & 0.1001\\
     REM&&0.958
&0.998&0.951 &0.975 && 0.2433\\
     LSP&&0.973
&1.0&0.968 & 0.984&& 0.1763\\
OPS&&0.932
&0.997&0.924 & 0.959 && 0.1701\\
    \bottomrule
    \end{tabular}
    \label{more_detection_poison_generation}}}
\hspace{1mm}
\parbox{.37\linewidth}{
\caption{Clean testing accuracy (\%) of models trained on the unlearnable CIFAR-10 dataset with different poisoning rate.  }
\centering
\scalebox{1.0}{
\setlength\tabcolsep{3.8pt}
\begin{tabular}{ c | c | c  c  c  c  }
\toprule
 {UEs} & Counter &0.2 & 0.4 & 0.6 & 0.8 \\
\midrule
\multirow{2}*{EM}& JPEG
& 85.03& 85.31 & 85.40& 85.31\\
& Ours& \textbf{93.50}
&\textbf{93.03}&\textbf{93.02} & \textbf{92.26}\\
\midrule
 \multirow{2}*{TAP}& JPEG&85.12
&85.60&84.92& 85.34\\
& Ours&\textbf{90.55}
&\textbf{90.78}&\textbf{90.93} & \textbf{91.10}\\
\midrule
\multirow{2}*{REM}& JPEG&84.64
&84.90& 84.62& 84.97\\
& Ours&\textbf{92.24}
&\textbf{92.51}&\textbf{92.23} & \textbf{90.86}\\
\midrule
\multirow{2}*{SEP}& JPEG&85.34
&85.22&85.06 & 85.06\\
& Ours& \textbf{90.86}
& \textbf{90.63} & \textbf{91.04} & \textbf{91.79}\\
 \midrule
 \multirow{2}*{LSP}& JPEG&85.22&85.34& 84.26 & 83.02\\
& Ours &\textbf{93.20}
&\textbf{92.85}&\textbf{92.16} & \textbf{92.16}\\
\midrule
\multirow{2}*{AR}& JPEG&85.31
&85.29& 85.33& 84.87 \\
& Ours&\textbf{92.77}
&\textbf{91.83}&\textbf{91.41} & \textbf{91.70}\\
 \midrule
 \multirow{2}*{OPS}& JPEG&85.12
&  84.89&  84.43& 83.01\\
& Ours&\textbf{93.15}
&\textbf{93.29}&\textbf{92.13} & \textbf{92.16}\\
\bottomrule
\end{tabular}
\label{more_partial}
}}
\vspace{-3mm}
\end{table}

\newpage
\section{Detailed implementation of the attack methods and competing defenses} \label{attack_methods}
As previous papers may have used varying code to generate perturbations and implemented defenses based on different codebases, we have re-implemented the majority of the attack and defensive methods by referencing their original code resources. In cases where the original paper did not provide code, we will specify the sources we used for implementation.
\subsection{Attack methods for generating UEs}
\textbf{NTGA.} For the implementation of NTGA {UEs}, we directly download the read-to-use unlearnable dataset from the official source of NTGA~\citep{ntga}.

\textbf{EM, TAP, and REM.} For the implementation of EM~\citep{em}, TAP~\citep{tap}, and REM~\citep{rem} {UEs}, we follow the official code of REM~\citep{rem}.

\textbf{SEP.} For the implementation of SEP~\citep{sep} {UEs}, we follow the official code of SEP~\citep{sep}.

\textbf{LSP.} For the implementation of LSP~\citep{lsp} {UEs}, we follow the official code of LSP~\citep{lsp}. Particularly, we set the patch size of the colorized blocks to 8 for both CIFAR-10, CIFAR-100, ImageNet-subset.

\textbf{AR.} For the implementation of AR {UEs}, we directly download the read-to-use unlearnable dataset from the official source of AR~\citep{ar}.

\textbf{OPS.} For the implementation of OPS.~\citep{ops} {UEs}, we follow the official code of OPS.~\citep{ops}. 

\begin{figure}[t]
\begin{minipage}{1.0\linewidth}
\centerline{{\includegraphics[width=0.8\linewidth]{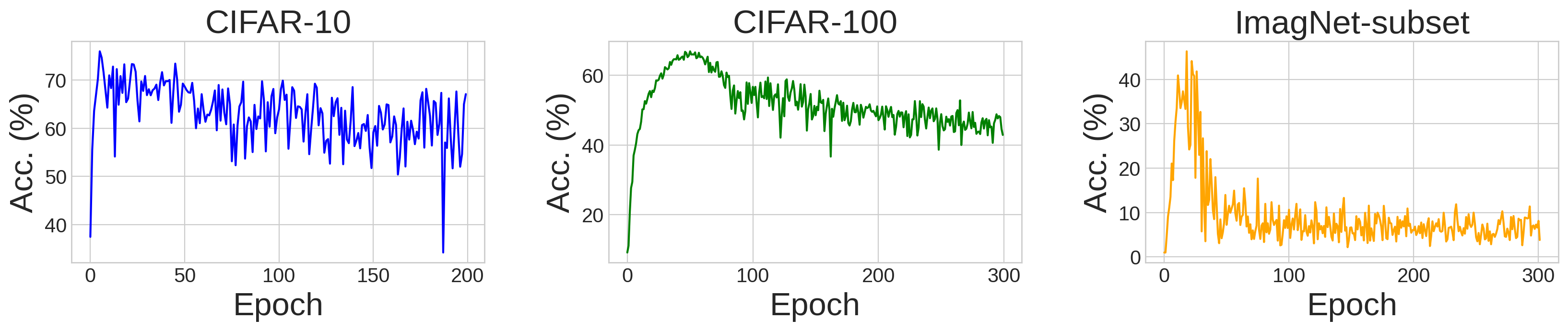}}}
\centerline{(a) EM~\citep{em}}
\vspace{2mm}
\end{minipage}
\begin{minipage}{1.0\linewidth}
\centerline{{\includegraphics[width=0.8\linewidth]{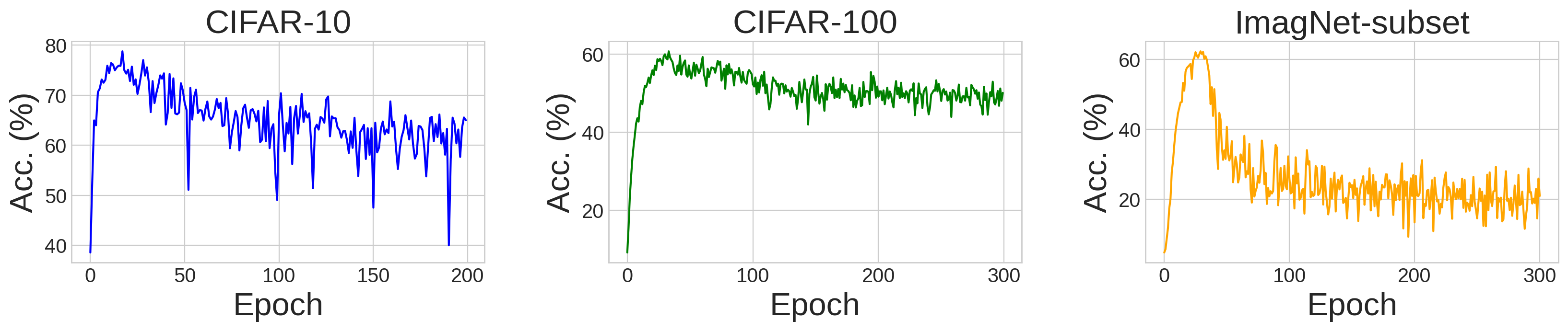}}}
\centerline{(b) REM~\citep{rem}}
\vspace{2mm}
\end{minipage}
\begin{minipage}{1.0\linewidth}
\centerline{{\includegraphics[width=0.8\linewidth]{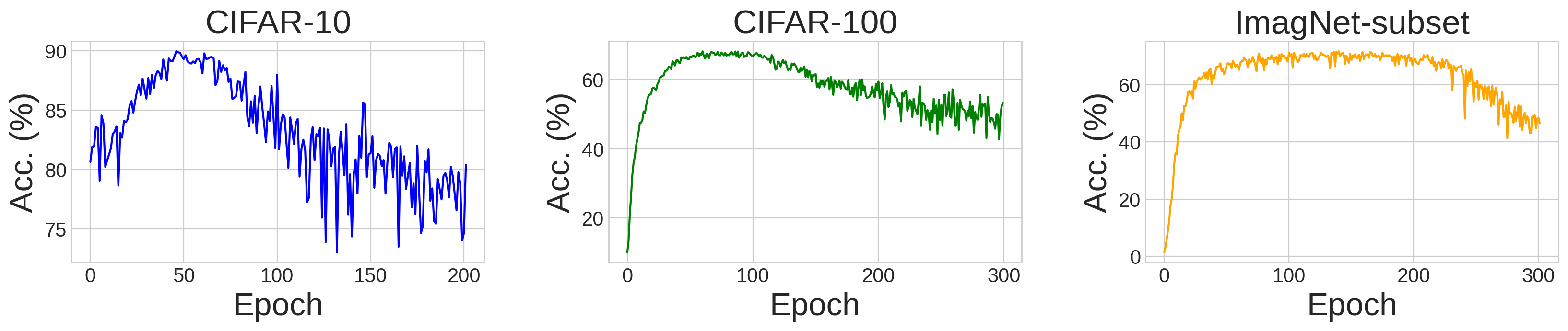}}}
\centerline{(c) LSP~\citep{lsp}}
\end{minipage}
    \caption{Test accuracy (\%) for each training epoch when using adversarial augmentation~\citep{qin2023learning}}
    \label{aa}
    \vspace{-4mm}
\end{figure}

\subsection{Competing defenses} \label{competing_defenses}
\textbf{Image shortcut squeezing (ISS).} For the implementation of ISS~\citep{iss}, which consists of bit depth reduction (depth decreased to 2), grayscale (using the official implementation by torchvision.transforms), JPEG compression (quality set to 10), we follow the official code of ISS~\citep{iss}. Although most of the reported results align closely with the original paper's findings, we observed that EM and REM {UEs} generated using the codebase of REM~\citep{rem} display a notable robustness to Grayscale, which differs somewhat from the results reported in the original paper.The unreported results for the performance of each compression on the CIFAR-100 and ImageNet datasets are presented in Table~\ref{iss_aa}.

\textbf{Adversarial training (AT).} For the implementation of adversarial training, we follow the official code of pgd-AT~\citep{pgd} with the adversarial perturbation subject to $\ell_{\infty}$ bound, and set $\epsilon=\frac{8}{255}$, iterations $T=10$, and step size $\alpha=\frac{1.6}{255}$. 

\textbf{AVATAR.} In our implementation of AVATAR, which employs a diffusion model trained on the clean CIFAR-10 dataset to purify unlearnable samples, we utilized the codebase from a benchmarking paper~\citep{qin2023apbench}. This choice was made since AVATAR~\citep{dolatabadi2023devil} does not offer official implementations.

\textbf{Adversarial augmentations (AA).} In our implementation of AA, we utilized the codebase from the original paper~\citep{qin2023learning}. AA comprises two stages. In the first stage, loss-maximizing augmentations are employed for training, with a default number of repeated samples set to K = 5. In the second stage, a lighter augmentation process is applied, with K = 1. In all experiments conducted on CIFAR-10, CIFAR-100, and the 100-class ImageNet subset, we strictly adhere to the same hyperparameters as detailed in the original paper. 
Nevertheless, we observed that this training-time method can partially restore the test accuracy if we report the highest accuracy achieved among all training epochs. However, it's worth noting that the model may still exhibit a tendency to overfit to the shortcut provided by the unlearnable samples. Consequently, this can lead to a substantial drop in test accuracy during the second stage, which employs lighter augmentations. The test accuracy for each training epoch is depicted in Figure~\ref{aa}.
Additionally, we have included the best accuracy for AA in Table~\ref{iss_aa}. It's notable that our results from the last epoch surpass the performance of AA, showcasing the superiority.

\begin{table}[t]
\parbox{.43\linewidth}{
  \centering
  \caption{Test acc. (\%) of models trained on CIFAR-10 {UEs}.  }
  \scalebox{1.0}{
  \setlength\tabcolsep{6.0pt}
    \centering
    \begin{tabular}{c | c | c  c  c}
    \toprule
     \multirow{2}*{Norm} & {Attacks}& w/o & AA & Ours  \\
     \cmidrule{2-5}
       & Clean& \textbf{94.57} & {92.66}& 93.29 \\
    \midrule
     \multirow{5}*{$\ell_{\infty} = 8$} &NTGA&11.10& 86.35& \textbf{89.21} \\
     &EM&12.26
& 76.00& \textbf{91.42} \\
     &TAP&25.44
& 71.56& \textbf{90.48} \\
     &REM&22.43
& 78.77&  \textbf{86.38} \\
     &SEP&6.63
& 71.95&  \textbf{90.74} \\
     \midrule
     \multirow{2}*{$\ell_{2} = 1.0$}& LSP&13.14& 89.97& \textbf{91.20} \\
     & AR&12.50
& 67.61&\textbf{91.77}\\
     \midrule
     {$\ell_{0} = 1$}& OPS&22.03
& 72.54& \textbf{88.95} \\
    \bottomrule
    \end{tabular}
}}
\hspace{1mm}
\parbox{.54\linewidth}{
  \centering
  \caption{Test acc. (\%) of models trained on CIFAR-100 {UEs}.  }
  \scalebox{1.0}{
  \setlength\tabcolsep{5.2pt}
    \centering
    \begin{tabular}{ c | c | c | c  c  c | c}
    \toprule
     {Attacks} & w/o & AA & BDR & Gray &JPEG & Ours \\
     \cmidrule{1-7}
       Clean & \textbf{77.61} & 70.22 & 63.52 & 71.59 & 57.85& 70.72\\
    \midrule
     EM&12.30
&66.84&61.91 & 48.83 & 58.08 & \textbf{68.79}\\
     TAP&13.44
&49.36&55.09 & 9.69 & 57.33 &\textbf{65.54}\\
     REM&16.80
&60.74&57.51 & 55.99 & 58.13 & \textbf{68.52}\\
     SEP&4.66
&37.73&31.95 & 4.47 & 57.76 & \textbf{64.02}\\
     \midrule
     LSP&2.91&68.22&22.13 & 44.18 & 53.06 &  \textbf{67.73}\\
     AR&2.71
&44.32&29.68 & 23.09 & 56.60 & \textbf{63.73}\\
     \midrule
     OPS&12.56
& 40.20 & 11.56 & 19.33 & 54.45 & \textbf{65.10}\\
    \bottomrule
    \end{tabular}
}}
\hspace{6.5mm}
\parbox{.9\linewidth}{
  \centering
  \caption{Test acc. (\%) of models trained on ImageNet subset {UEs}.  }
    \centering
     \scalebox{1.0}{
     \setlength\tabcolsep{3.5pt}
    \begin{tabular}{c | c | c | c  c  c | c}
    \toprule
     {Attacks} & w/o & AA & BDR & Gray &JPEG & Ours \\
     \cmidrule{1-7}
       Clean & \textbf{80.52} & 73.66 & 75.84 & 76.92 & 72.90 & 76.78\\
    \midrule
     EM&1.08
&46.30&2.78 & 14.02 & 72.44 & \textbf{74.80}\\
     TAP&12.56
&72.10&45.74 & 33.66 & 73.24 & \textbf{76.56}\\
     REM&2.54
&62.30&57.51 & 55.99 & 58.13 & \textbf{72.56}\\
     \midrule
     LSP&2.50&71.72&22.13 & 44.18 & 53.06 &  \textbf{76.06}\\
    \bottomrule
    \end{tabular}
}}
\label{iss_aa}
\end{table}

\clearpage
\newpage
\section{Visual Results}\label{more_visual_results}
\begin{figure*}[t]
\centering
\includegraphics[width=1.0\linewidth]{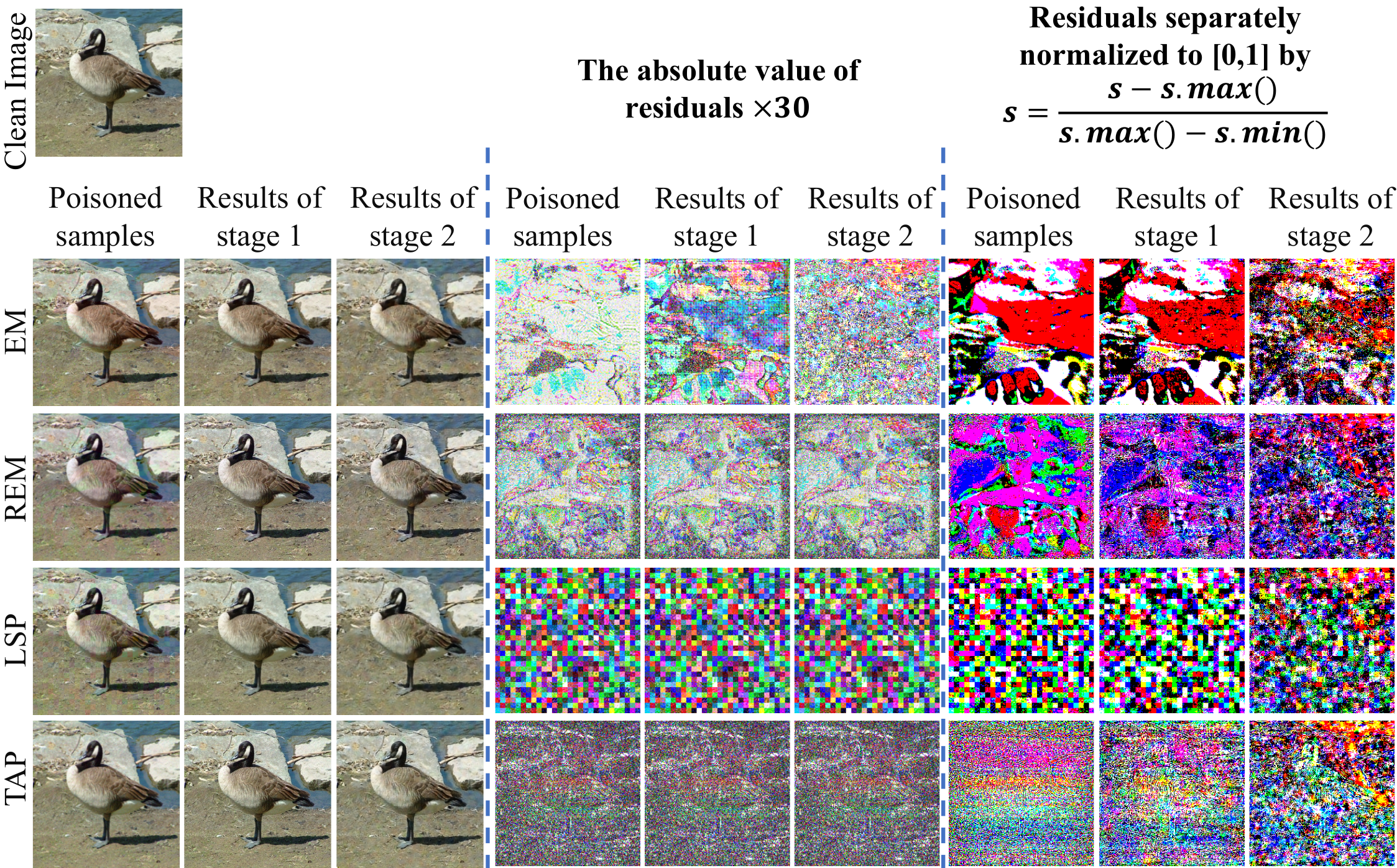}
\vspace{-4.5mm}
\caption{
Visual results of images before/after purification. Results of stage 2 denote the final purified results. The image is from ImageNet-subset, and the residuals to the clean images are normalized by two ways.
}
\label{imagenet_examples}
\end{figure*}

In this section, we present visual results of the purification process on the ImageNet-subset.
As depicted in Figure~\ref{imagenet_examples}, the purification carried out during stage 1 is effective in removing a significant portion of perturbations, particularly for LSP {UEs}. The remaining {perturbations} are subsequently eliminated in stage 2, resulting in completely poison-free data.

\clearpage
\section{{Comparison with non-variational auto-encoders}\label{non-variational}}
\begin{figure*}[t]
\centering
\includegraphics[width=1.0\linewidth]{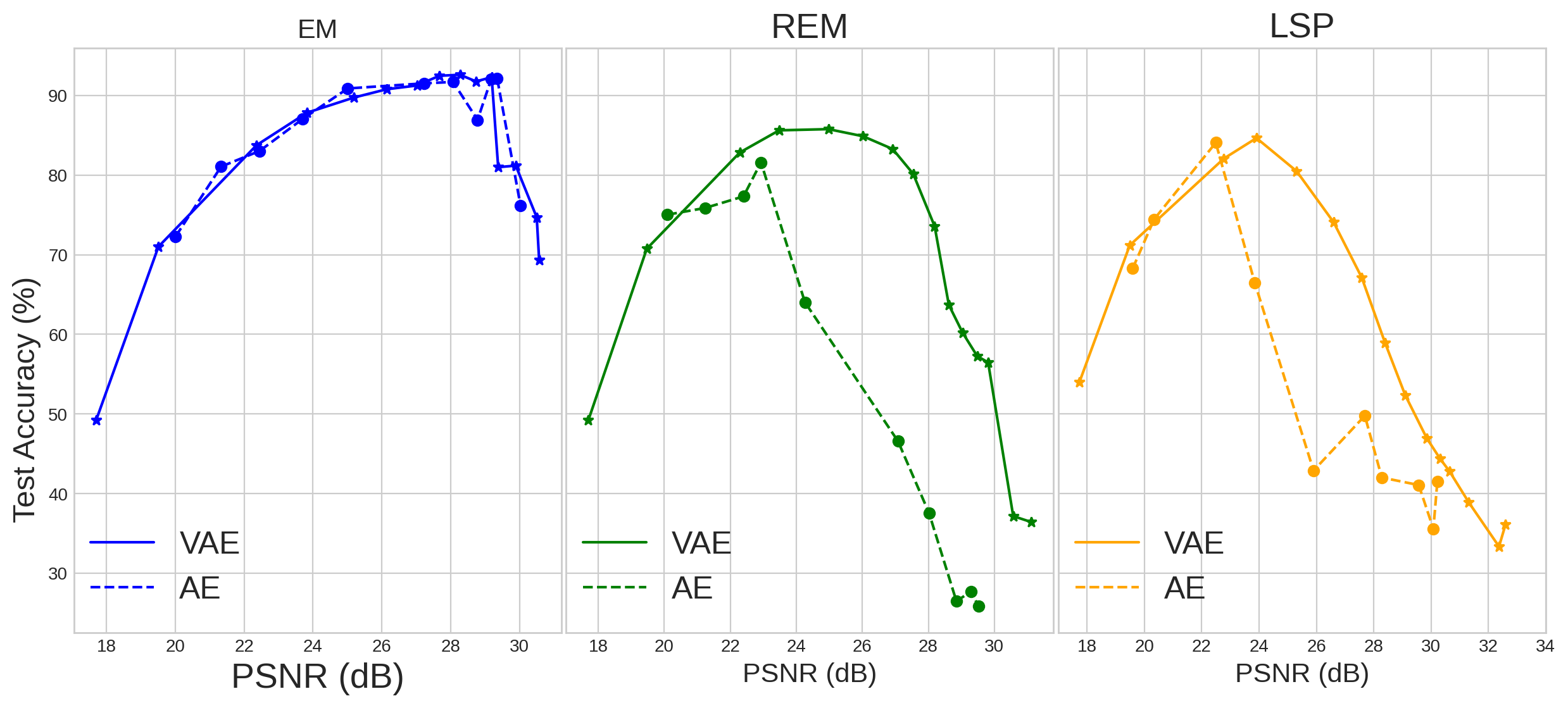}
\vspace{-4mm}
\caption{
Comparison between VAEs and AEs: PSNR Vs. Test Acc. Specifically, we include EM, REM, and LSP as attack methods here.
}
\label{comparison_with_AE}
\end{figure*}

{
In this section, we conduct experiments on purification using non-variational auto-encoders (AEs) with an information bottleneck. To achieve non-variational auto-encoders with different bottleneck levels, we modify the width of the features within the auto-encoder architecture. This results in models with varying parameter numbers. 
}
{
Then, we proceed to train the AE on the unlearnable CIFAR-10 dataset, and test on the clean test dataset with classifiers trained on the purified dataset. 
As depicted in Figure~\ref{comparison_with_AE}, when considering the similar level of reconstruction quality measured by PSNR, VAEs exhibit a greater capacity to remove perturbations in both the REM and LSP UEs. However, for EM UEs, the outcomes are comparable. These observations align with the theoretical analysis presented in Section 3.3.
}

\newpage
\begin{table}[t]
    \centering
\setlength\tabcolsep{5.5pt}
    \caption{Computation requirement of the proposed methods. }
    \centering
    \scalebox{1.0}{
    \begin{tabular}{ c | c  c  c | c  }
    \toprule
     Component & \shortstack{Train D-VAE \\for twice} &\shortstack{Perform inference on the\\ unlearnable data three times}&\shortstack{Train a \\classifier}&Total Time\\
    \midrule
     Our method &23 minutes	&less than 2 minutes&16 minutes&41 minutes\\
     Adversarial Training&N.A.&N.A.&229 minutes&229 minutes\\
    \bottomrule
    \end{tabular}
    \label{computation}}
\end{table}
\section{{Computation and Comparison with JPEG compression}\label{computation_cost}}
{
In this section, we present the computation requirement and the compassion with JPEG compression.
The Table~\ref{computation} below presents the training time for D-VAE, the inference time for the unlearnable dataset, and the time to train a classifier using the purified dataset. For comparison, we include the training-time defense \textbf{Adversarial Training}. It's important to note that the times are recorded using CIFAR-10 as the dataset, PyTorch as the platform, and a single Nvidia RTX 3090 as the GPU. As can see from the results, the total purification time is approximately one and a half times longer than training a classifier, which is acceptable. Compared to adversarial training, our methods are about 5 times faster. Additionally, our method achieves an average performance around 90\%, which is 15\% higher than the performance achieved by adversarial training.
}

{
We also note a limitation in the JPEG compression approach used in ISS~\citep{iss}—specifically, they set the JPEG quality to 10 to purify unlearnable samples, resulting in significant image degradation. In the Table~\ref{jpeg_other}, we present results using JPEG with various quality settings. Notably, our proposed methods consistently outperform JPEG compression when applied at a similar level of image corruption. Therefore, in the presence of larger perturbation bounds, JPEG may exhibit suboptimal performance. Moreover, our method excels in eliminating the majority of perturbations in the first stage, rendering it more robust to larger perturbation bounds. Table 5~\ref{larger_perturbation} of the main paper illustrates that when confronted with LSP attacks with larger bounds, our method demonstrates significantly smaller performance degradation compared to JPEG (with quality 10), \textit{e.g.,} 86.13 Vs. 41.41 in terms of test accuracy.
}

\begin{table}[t]
    \centering
\setlength\tabcolsep{2.3pt}
    \caption{Results using JPEG with various quality settings. The experiments are on CIFAR-10 dataset.}
    \centering
    \scalebox{1.0}{
\begin{tabular}{c|cccc|c}
\toprule
\shortstack{Defenses\\/Attacks} & \shortstack{JPEG (quality 10)\\PSNR 22} & \shortstack{JPEG (quality 30)\\PSNR 25} & \shortstack{JPEG (quality 50)\\PSNR 27} & \shortstack{JPEG (quality 70)\\PSNR 28}& \shortstack{Ours\\PSNR 28}\\ 
\midrule
NTGA & 78.97 & 66.83 & 64.28 & 60.19 & \textbf{89.21} \\ 
EM & 85.61 & 70.48 & 54.22 & 42.23 & \textbf{91.42}\\ 
TAP & 84.99 & 84.82 & 77.98 & 57.45 & \textbf{90.48} \\ 
REM & 84.40 & 77.73 & 71.19 & 63.39 & \textbf{86.38} \\ 
SEP & 84.97 & 87.57 & 82.25 & 59.09 & \textbf{90.74} \\ 
LSP & 79.91 & 42.11 & 33.99 & 29.19 & \textbf{91.20} \\ 
AR & 84.97 & 89.17 & 86.11 & 80.01 & \textbf{91.77} \\ 
OPS & 77.33 & 79.01 & 68.68 & 59.81 & \textbf{88.96} \\ 
\midrule
Mean & 78.89 & 74.71 & 67.33 & 56.42 & \textbf{90.02} \\ 
\bottomrule
\end{tabular}
    \label{jpeg_other}}
\end{table}

\newpage
\section{{More experiments on training D-VAE on attack methods with various target values on the KLD Loss}}
{
Some concerns regarding whether a sizable component of the perturbation will end up being learned into $\boldsymbol{\hat{x}}$ may arise in certain cases, such as when the target value on the KLD loss is not set low. Nevertheless, when the KLD loss is set to a low value, the presence of perturbations in the reconstructed $\boldsymbol{\hat{x}}$ is shown to be minimal. This observation is supported by both empirical experiments in Section 3.2 and theoretical explanations provided in Section 3.3. These outcomes are primarily attributed to the fact that the reconstruction of $\boldsymbol{\hat{x}}$ depends on the information encoded in the latent representation $\boldsymbol{z}$, \textit{i.e.,} $\boldsymbol{\hat{x}}$ is directly generated from $\boldsymbol{z}$ using a decoder. The theoretical insights discussed in Section 3.3 highlight that \textbf{Theorem 1} indicates that perturbations which create strong attacks tend to have a larger inter-class distance and a smaller intra-class variance. Additionally, \textbf{Theorem 2} and \textbf{Remark 1} indicate that perturbations possessing these characteristics are more likely to be  eliminated when aligning the features with a normal Gaussian  distribution (as done by the VAE).  
}

\begin{figure}[t]
\begin{minipage}{0.48\linewidth}
\centerline{{\includegraphics[width=0.9\linewidth]{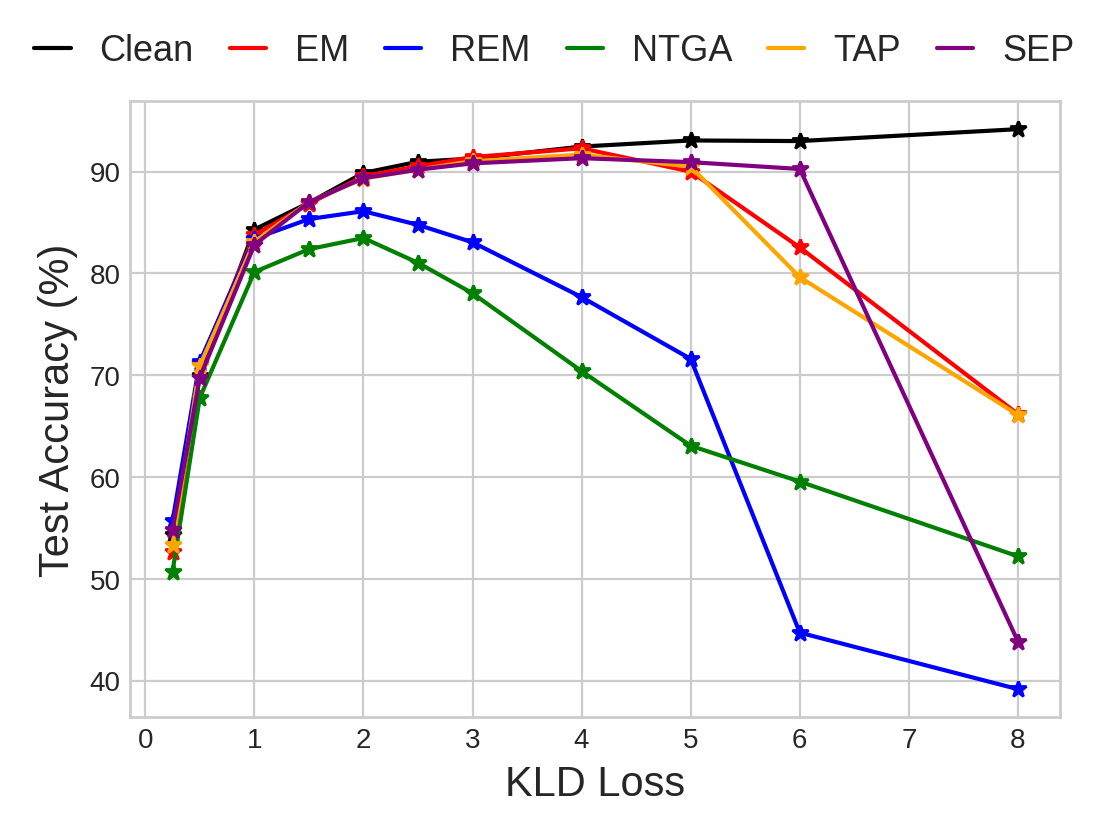}}}
\centerline{(a) $\ell_{\infty}=\frac{8}{255}$}
\end{minipage}
\begin{minipage}{0.48\linewidth}
\centerline{{\includegraphics[width=0.9\linewidth]{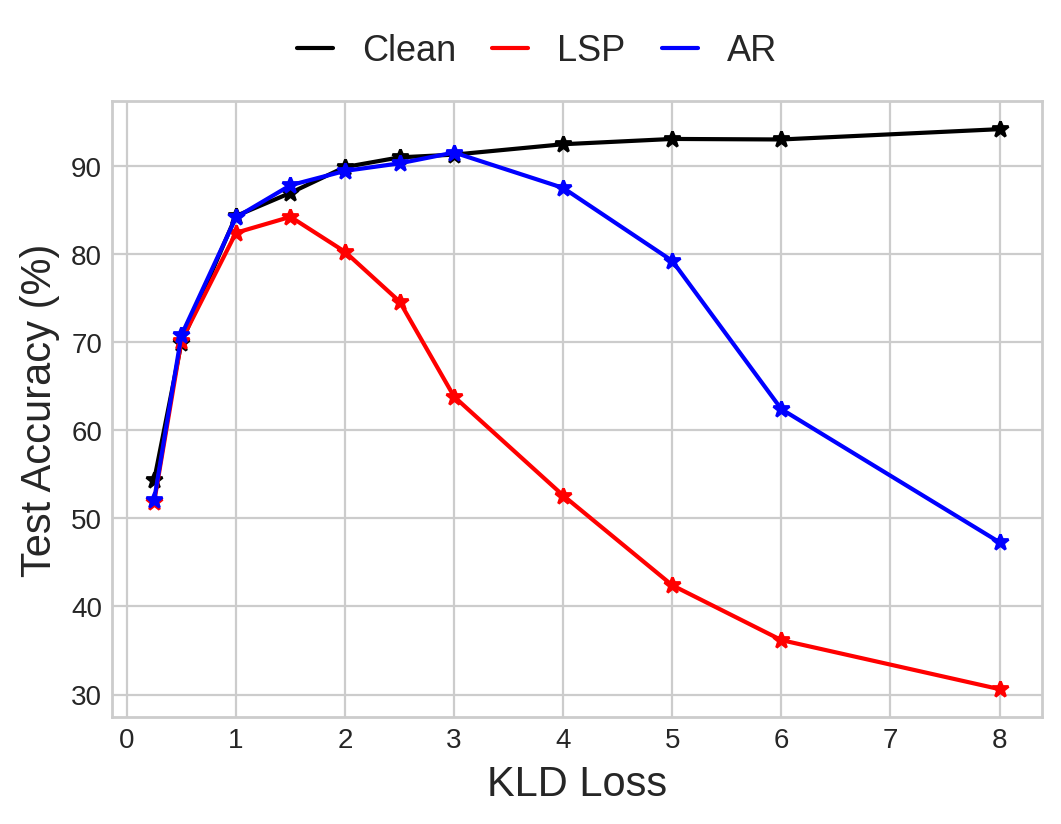}}}
\centerline{(b) $\ell_2=1.0$}
\end{minipage}
\begin{minipage}{0.48\linewidth}
\centerline{{\includegraphics[width=0.9\linewidth]{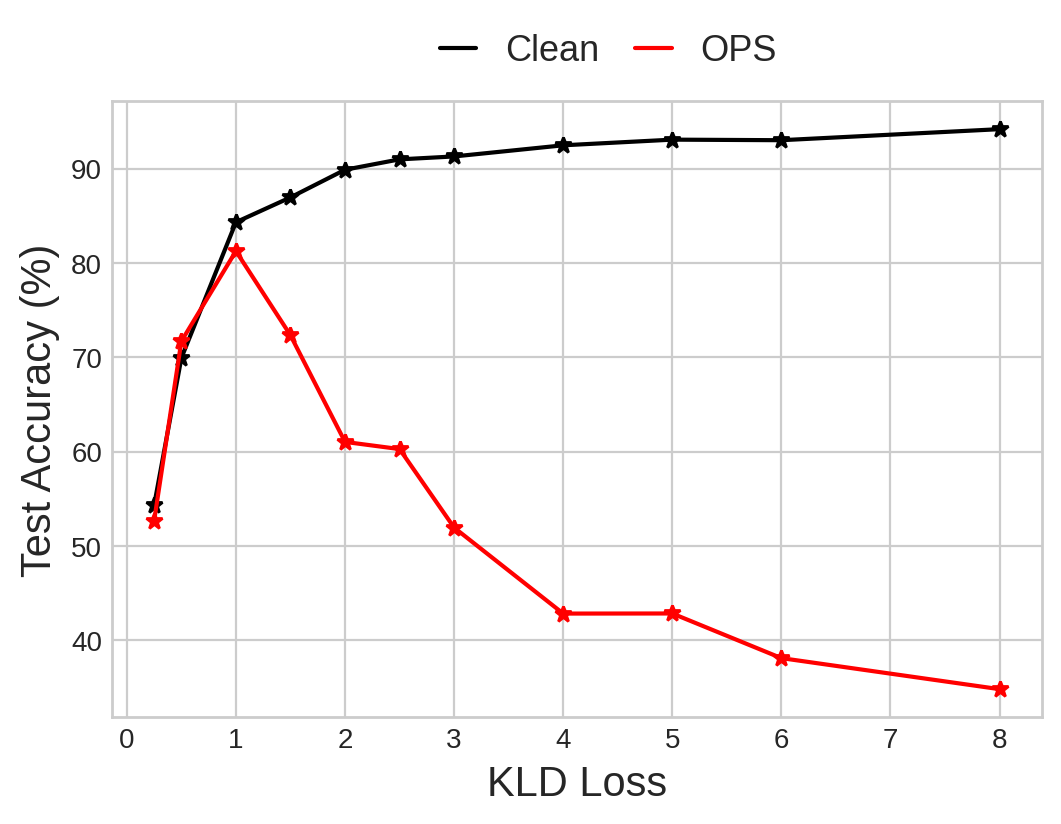}}}
\centerline{(c) $\ell_0=1$}
\end{minipage}
\hspace{5mm}
\begin{minipage}{0.48\linewidth}
\centerline{{\includegraphics[width=0.9\linewidth]{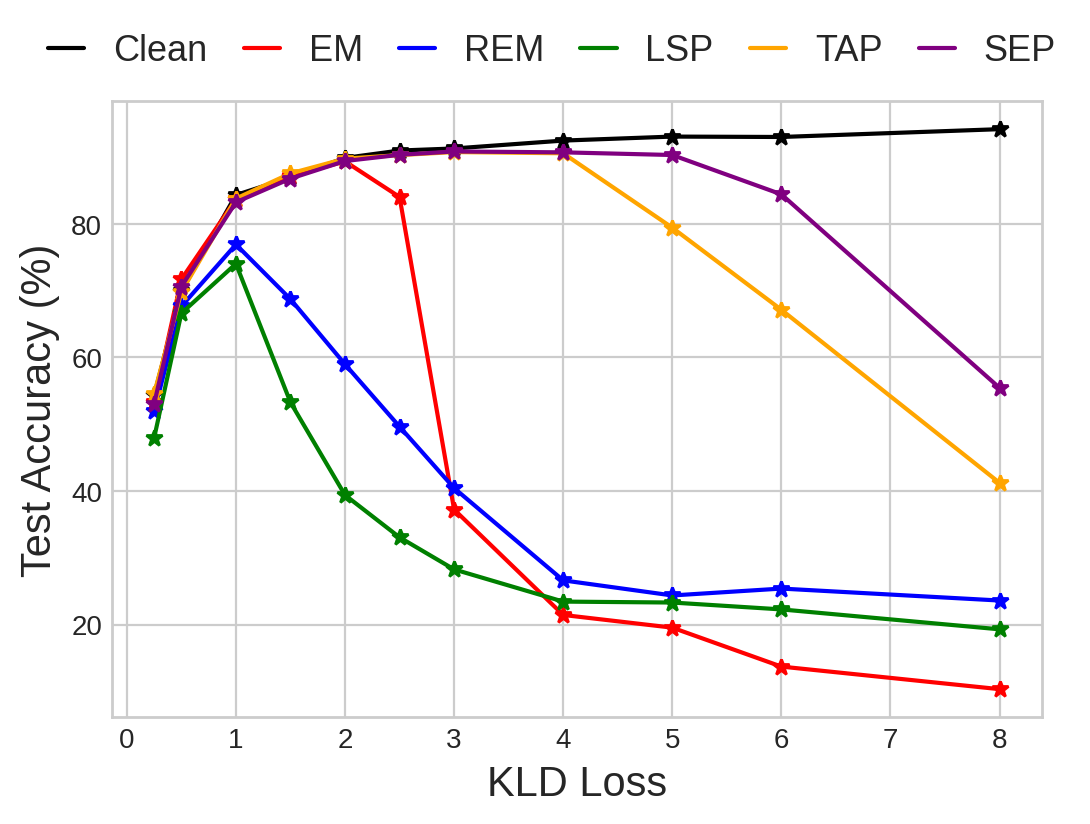}}}
\centerline{(d) $\ell_{\infty}=\frac{16}{255}$ or $\ell_2=2.0$}
\end{minipage}
\vspace{-2mm}
    \caption{Results using D-VAEs: Test Acc. Vs. KLD Loss is assessed on the unlearnable CIFAR-10.}
    \label{dvae_acc_kld}
\end{figure}

{
To further validate these observations, we now include additional experiments in Appendix G by training D-VAE on all attack methods with various target values for the KLD loss. Additionally, we have performed experiments on attacks with larger perturbations. Notably, we have added results on the clean dataset for comparison. As depicted in Figure~\ref{dvae_acc_kld}, when the target value on the KLD loss is set below 1.0, the curves of the results on the unlearnable dataset align closely with the results on the clean dataset. Furthermore, as the target value decreases, the removal of perturbations in the reconstructed $\boldsymbol{\hat{x}}$ increases. While it is evident that larger perturbations may be better retained in $\boldsymbol{\hat{x}}$, it is a cat-and-mouse game between defense and attack. Additionally, larger perturbations tend to be more noticeable. These findings affirm that the observations hold for all existing attack methods, and setting a low target value (\textit{e.g.,} 1.0, as in the main experiments) on the KLD loss significantly ensures that $\boldsymbol{\hat{x}}$ contains few perturbations.
}

\section{Selection of various $\boldsymbol{kld_1, kld_2}$.}
\label{Selection_of_various_kld}
To showcase that our method is tolerant to the selection of $kld_1$ and $kld_2$, we conduct experiments on the CIFAR-10 dataset. We present the defensive performance against eight UEs methods as shown the Table~\ref{choice_kld}. We denote our method with different hyperparameters as "Ours ($kld_1/kld_2$)". Our findings indicate that while varying hyperparameters may result in only a slight decrease in the effectiveness of our proposed method (less than 1\%).
Furthermore, as depicted in Table~\ref{larger_perturbation} in our paper, when confronted with UEs with larger perturbation bounds, our method with the exact same $kld_1$ and $kld_2$ values exhibits slight performance degradation, yet still manages to achieve superior performance overall.

\begin{table}[t]
\centering
\setlength\tabcolsep{4.0pt}
\caption{Comparison of Defenses}
\label{defenses_comparison}
\begin{tabular}{l|c|c|c|c|c|c|c|}
\toprule
\textbf{Defenses} & \textbf{JPEG} & \textbf{AVA.} & \textbf{Ours ($1.5/2.5$)} & \textbf{Ours ($1.5/3.0$)} & \textbf{Ours ($0.5/3.0$)} & \textbf{Ours ($1.0/3.5$)} & \textbf{Ours ($1.0/3.0$) as reported} \\ \midrule
NTGA & 78.97 & 80.72 & 87.18 & 87.64 & 89.18 & 88.65 & 89.21 \\ \midrule
EM & 85.61 & 89.54 & 90.65 & 91.14 & 91.64 & 91.86 & 91.42 \\ \midrule
TAP & 84.99 & 89.13 & 90.60 & 90.98 & 90.38 & 91.52 & 90.48 \\ \midrule
REM & 84.40 & 86.06 & 86.60 & 85.77 & 85.47 & 84.58 & 86.38 \\ \midrule
SEP & 84.97 & 89.56 & 90.02 & 90.76 & 90.23 & 91.31 & 90.74 \\ \midrule
LSP & 79.91 & 81.15 & 89.61 & 90.50 & 91.40 & 91.72 & 91.20 \\ \midrule
AR & 84.97 & 89.64 & 90.23 & 91.29 & 90.80 & 90.52 & 91.77 \\ \midrule
OPS & 77.33 & 71.62 & 87.89 & 86.18 & 89.39 & 86.50 & 88.96 \\ \midrule
Mean & 82.64 & 84.67 & 89.09 & 89.30 & 89.56 & 89.58 & \textbf{90.02} \\ \bottomrule
\end{tabular}
\label{choice_kld}
\end{table}

\end{document}